# iMaNGA: mock MaNGA galaxies based on IllustrisTNG and MaStar SSPs – II. The catalogue

Lorenza Nanni[1]★ Daniel Thomas[1,2] James Trayford[1] Claudia Maraston[1] Justus Neumann[3] David R. Law[4] Lewis Hill[1] Annalisa Pillepich[3] Renbin Yan[5] Yanping Chen[6] and Dan Lazarz[7]

[1]*Institute of Cosmology & Gravitation, University of Portsmouth, Dennis Sciama Building, Portsmouth PO1 3FX, UK*
[2]*School of Mathematics and Physics, University of Portsmouth, Lion Gate Building, Portsmouth PO1 3HF, UK*
[3]*Max-Planck-Institut fur Astronomie, Konigstuhl 17, D-69117 Heidelberg, Germany*
[4]*Space Telescope Science Institute, 3700 San Martin Drive, Baltimore, MD 21218, USA*
[5]*Department of Physics, The Chinese University of Hong Kong, Shatin, N.T., Hong Kong, SAR 999077, China*
[6]*New York University Abu Dhabi, Abu Dhabi PO Box 129188, United Arab Emirates*
[7]*Department of Physics and Astronomy, University of Kentucky, 505 Rose St., Lexington, KY 40506-0057, USA*

**ABSTRACT**
Strengthening the synergy between simulations and observations is essential to test galaxy formation and evolution theories. To achieve this goal, in the first paper of this series, we presented a method to generate mock SDSS-IV/Mapping Nearby Galaxies at Apache Point Observatory (MaNGA) integral-field spectroscopic galaxy observations from cosmological simulations. In this second paper, we build the iMaNGA catalogue consisting of ∼1000 unique galaxies from the TNG50 cosmological simulations, selected to mimic the SDSS-IV/MaNGA-Primary sample selection. Here, we present and discuss the iMaNGA sample and its comparison to the MaNGA Primary catalogue. The iMaNGA sample well recovers the MaNGA-Primary sample in terms of stellar mass versus angular size relation and spatial resolution. The Sérsic index versus angular size relation, instead, is not reproduced well by the simulations, mostly because of a paucity of high-mass elliptical galaxies in TNG50. We also investigate our ability to recover the galaxy kinematics and stellar population properties with full-spectral fitting. We demonstrate that 'intrinsic' and 'recovered' stellar kinematics, stellar ages, and metallicities are consistent, with residuals compatible with zero within $1\sigma$. Also 'intrinsic' and 'recovered' star formation histories display a great resemblance. We conclude that our mock generation and spectral fitting processes do not distort the 'intrinsic' galaxy properties. Therefore, in the third paper of this series, we can meaningfully test the cosmological simulations, comparing the stellar population properties and kinematics of the iMaNGA mock galaxies and the MaNGA observational results.

## 1 INTRODUCTION

During cosmic history, galaxies are shaped by complex physics acting on multiple scales (Somerville & Davé 2015). Hence, hydrodynamical simulations in a cosmological context have been utilized to theoretically predict what we observe (Vogelsberger et al. 2020). Nowadays, large-scale hydrodynamical simulations of galaxy formation are available, such as Illustris (Nelson et al. 2015), IllustrisTNG (Nelson et al. 2019a), and EAGLE (Schaye et al. 2014): in large cosmological volumes, baryonic matter, and dark matter evolve together from the primordial density fluctuations to the local universe. Thanks to these large simulated samples, we can now test theoretical predictions against the tremendous amount of observational data provided by modern surveys, e.g. the Cosmic Assembly Near-infrared Deep Extragalactic Legacy Survey (CANDLES; Grogin et al. 2011), the *Sloan Digital Sky Surveys* (*SDSS*; York et al. 2000; Abazajian et al. 2003), the Calar Alto Legacy Integral Field Area survey (CALIFA; Sánchez et al. 2012), the Sydney-AAO Multi-object Integral field spectrograph (SAMI; Allen 2014), and Mapping Nearby Galaxies at Apache Point Observatory (MaNGA; Bundy et al. 2015).

'Forward modelling' is a technique to compare theory to observations, which places model galaxies on the observational plane taking into account a variety of observational effects. This has been employed in e.g. Tonini et al. (2010), Snyder et al. (2015), Torrey et al. (2015), Trayford et al. (2015, 2017), Bottrell et al. (2017), Rodriguez-Gomez et al. (2019), Huertas-Company et al. (2019), and Schulz et al. (2020). Below we provide a brief synopsis of each of these works. Tonini et al. (2010) and Henriques et al. (2012) show that semi-analytical models of galaxy formation (for an overview of these models see Vogelsberger et al. 2020) better reproduce the observed colours and near-infrared luminosities of high-redshift ($z \sim 2–3$) massive galaxies when calculated with stellar population models accounting for the thermally pulsing asymptotic

★ E-mail: lorenza.nanni@port.ac.uk (LN) daniel.thomas@port.ac.uk (DT) claudia.maraston@port.ac.uk (CM)

giant branch (TP-AGB) phase of stellar evolution. Snyder et al. (2015) include the effect of a Gaussian Point Spread Function and noise into synthetic images from Illustris galaxies in order to study how optical galaxy morphology depends on mass and star formation rate in the simulations. Torrey et al. (2015) develop a method to build a catalogue of 7000 synthetic images and 40 000 integrated spectra from the Illustris simulations at redshift 0, proving how, from the synthetic data products, it is possible to produce monochromatic or colour-composite images, perform SED fitting, classify morphology, and determine galaxy structural properties, as for the analysis of real galaxies. Trayford et al. (2015) include the effects of obscuration by dust in birth clouds and the interstellar medium in EAGLE simulated galaxies, using a two-component screen model. In Trayford et al. (2017), the dust effect is included with radiative transfer simulations, demonstrating an improvement between the predicted optical colours as a function of the stellar mass with the observed ones. Bottrell et al. (2017) generate synthetic images from the Illustris simulations, including noise and the effect of the point spread function, in order to carry out bulge + disc decompositions for *SDSS*-type galaxy images. Their work reveals that galaxies in Illustris are approximately twice as large and 0.7 mag brighter on average than galaxies in the *SDSS*, because of a significant deficit of bulge-dominated galaxies in Illustris for $\log M_*/M_\odot < 11$. Rodriguez-Gomez et al. (2019) generate synthetic images of ∼27 000 galaxies from the IllustrisTNG and Illustris, to match Pan-STARRS (Chambers et al. 2016) galaxy observations. The synthetic and real Pan-STARRS images are analysed with the same code (STATMORPH). The comparison reveals that the optical morphologies of IllustrisTNG galaxies are in good agreement with observations, improving the predictions of the original Illustris simulation. However, the IllustrisTNG model still does not reproduce the observed strong morphology–colour relation because of an excess of both red discs and blue spheroids. Moreover, at a fixed stellar mass, observations find discs to be larger than spheroids, while IllustrisTNG does not predict this trend. Huertas-Company et al. (2019) select around 12 000 galaxies in TNG100 to generate mock *SDSS* images, using the radiative transfer code SKIRT (Camps & Baes 2014; Baes & Camps 2015) and including PSF and noise to mock *SDSS* *r*-band images. Observed and model morphologies are studied with a Convolutional Neural Network. The mass–size relations of the galaxies, divided by morphological type, match satisfactorily. However, there are discrepancies at the high-mass end of the stellar mass functions (SMF), which is dominated by disc galaxies in TNG100 and by early-type galaxies in *SDSS*. Schulz et al. (2020) investigate the relationship between the UV slope, β, and the ratio between the infrared and UV luminosities (IRX) of galaxies from TNG50 on 7280 star-forming main-sequence (SFMS) galaxies. A general good agreement is found at $z \geq 1$. However, they find a redshift-dependent systematic offset concerning empirically derived local relations, with the TNG50 IRX-β relation shifting towards lower β and steepening at higher redshifts. This selection of papers highlights how complex the comparison between observations and simulations is and the need for including observational effects in the simulations in order to allow for a close comparison.

This is the approach we take in the papers of this series, which focus on simulating the MaNGA sample, which is an integral-field spectroscopic survey of 10 010 nearby galaxies (see Section 2.3). In Nanni et al. (2022), hereafter Paper I, we introduced our forward modelling procedure to generate realistic mock MaNGA-like galaxies. The main novelties of our method are: the adoption of the MaStar stellar population models (Maraston et al. 2020)[1] that are based on stellar spectra obtained with the same MaNGA spectrograph (see Section 2.2 for details); a radiative transfer-based treatment of the dust; the reconstruction of a wavelength-dependent spectral noise based on MaNGA data; the use of the MaNGA effective point spread function to include observational effects such as dithering. Furthermore, we follow the steps of the MaNGA Data Analysis Pipeline (DAP; Westfall et al. 2019) to analyse the mock data. Specifically, we use two spectral fitting algorithms, namely PPXF (see Cappellari 2017) in order to obtain stellar and gas kinematics and FIREFLY (Wilkinson et al. 2017) to obtain the stellar populations' properties – age, chemical composition, star formation history (SFH), reddening, and stellar and remnant masses – as in several analysis of the MaNGA data (e.g. Goddard et al. 2016; Goddard et al. 2017; Neumann et al. 2021, 2022). As for cosmological simulations of galaxy formation and evolution here we adopt IllustrisTNG (Pillepich et al. 2018b; Nelson et al. 2019a), but we stress that our procedure can be easily applied to any other simulation suite.

In this paper, we describe how we construct a mock MaNGA-like catalogue – which we call the 'iMaNGA sample' – by applying the MaNGA-Primary target selection boundaries in redshift and *i*-band absolute magnitude (see Section 2.3) to the TNG50 and employing the post-processing and analysis pipeline presented in Paper I over this selection. This results in ∼1000 unique TNG50 galaxies obeying the selection. Here, we present and discuss the general properties of the mock galaxy catalogue, i.e. morphology, kinematics, and stellar populations. We then discuss how iMaNGA compares to the MaNGA-Primary sample, in particular focusing on the mass versus angular size relation, the spatial resolution and the Sérsic index versus angular size and mass relations, see Section 6. We finally demonstrate our ability to recover the truth values, i.e. the 'intrinsic' galaxy properties in the simulations. In the third paper of this series, we shall conduct a systematic comparison between our mock galaxies and observational results, including our own recent analysis published in Neumann et al. (2021).

The paper is organized as follows. Data and models in use are described in Section 2, while our forward modelling procedure is recalled in Section 3. The construction of the mock galaxy catalogue is presented in Section 4 and results are discussed in Section 5. In particular, we show the general properties of the iMaNGA sample in Section 5.1; we illustrate the morphological characteristics of the iMaNGA sample in Section 5.2; we compare the MaNGA-Primary sample to the iMaNGA one in Section 5.3; we present the results of the analysis of the kinematics in Section 5.4; we study the stellar population properties in Section 5.5. Also, in Section 6 we discuss other works on the construction of MaNGA-like galaxies from simulations. We draw our conclusions in Section 7.

## 2 INPUT MODELS AND DATA

Here, we recap the description of models and data used in this work.

### 2.1 The IllustrisTNG simulation suite

IllustrisTNG (Marinacci et al. 2018; Naiman et al. 2018; Nelson et al. 2018, 2019a, b; Pillepich et al. 2018b, 2019; Springel et al. 2018) is a suite of large-scale hydrodynamical simulations of galaxy formation and evolution, based on its predecessor Illustris (Genel et al. 2014; Vogelsberger et al. 2014; Sijacki et al. 2015). IllustrisTNG, while maintaining the fundamental approach and physical models of Illustris, expands its scientific goal with larger volumes (up to 300 Mpc instead of 100 Mpc), and higher resolution (up to a mass resolution for the baryonic matter of $8.5 \times 10^4$ M$_\odot$ instead of $1.6 \times 10^6$ M$_\odot$). Moreover, new physics is incorporated (including magnetic fields, and dual-mode black hole feedback, as described in Weinberger et al. 2017; Pillepich et al. 2018a). The fundamental physical processes comprised in these projects are the formation of cold dense gas clouds and stars; the stellar populations' evolution and stellar wind and feedback; the supernovae physics and evolution; the formation of supermassive BHs and their accretion, radiation, and feedback; the interstellar medium and its chemical enrichment. Indeed, the formation and evolution of galaxies are shaped by these processes which act across a broad range of spatial and time-scales, governing galaxies' fundamental characteristics, such as their stellar and gas content, star formation activity, chemical composition, morphology, and also their interactions with the external environment, e.g. in a cluster. Star formation in particular occurs stochastically when the gas number density is $\geq 0.13$ particle/cm$^{-3}$ according to a Chabrier (2003) initial mass function (IMF) and assuming the Kennicutt–Schmidt law (Schmidt 1959; Kennicutt 1989).

---

[1] https://www.icg.port.ac.uk/mastar/

IllustrisTNG simulates three physical box sizes, with cubic volumes of roughly 50, 100, and 300 Mpc side lengths (named TNG50, TNG100, and TNG300, respectively). Each run has a different resolution. Particularly, in TNG50 (Nelson et al. 2019b; Pillepich et al. 2019), the gravitational softening for baryonic and dark matter is: $\epsilon_{\rm gas,min} = 74$ pc and $\epsilon_{\rm DM,min} = 288$ pc; the mass resolution for baryonic and dark matter is: $m_{\rm bar} = 8.5 \times 10^4$ M$_\odot$ and $m_{\rm DM} = 4.5 \times 10^5$ M$_\odot$.[2] Each run outputs 100 snapshots from redshift 20.05 to redshift 0.0. Haloes and subhaloes are identified with the Friends-of-Friends and the SUBFIND algorithms, respectively (see Springel et al. 2001b; Nelson et al. 2015).

In this paper, we focus on subhaloes simulated by TNG50 and identified in snapshots from redshift 0.15 to redshift 0.01, which approximately corresponds to the redshift range observed with MaNGA (see Section 2.3). TNG50 is chosen because it allows the high spatial resolution typical of the MaNGA datacubes (pixel size of 0.5 arcsec, i.e. a spatial sampling raging from $\approx 100$ pc at $z \approx 0.01$ to $\approx 1.5$ kpc at $z \approx 0.15$, Section 2.3). A further discussion about the subhalo selection is presented in Section 4.

### 2.2 MaStar: SDSS-based stellar population models

We use stellar population models from Maraston et al. (2020) which adopt the MaNGA stellar library MaStar (Yan et al. 2019) for the definition of stellar spectra as a function of effective temperature, gravity, and chemical composition in the population synthesis.[3] MaStar (Abdurro'uf et al. 2022) consisting of $\sim 60\,000$ is the largest stellar library ever assembled. MaStar stellar spectra were obtained with MaNGA fiber bundles and the BOSS optical spectrographs, i.e. the same observational set-up as for MaNGA galaxy observations (see Section 2.3). Therefore, the stellar spectra and the correspondent population models share the same wavelength range, spectral resolution and flux calibration as the MaNGA datacubes.

---

[2]In all simulations, the Planck cosmology from Ade et al. (2016) is adopted, with matter density parameter $\Omega_{\rm m} = 0.31$; dark energy density parameter $\Omega_\Delta = 0.69$; Hubble constant $H_0 = 100$ h km s$^{-1}$ Mpc, with $h = 0.68$; matter power spectrum amplitude of $\sigma_8 = 0.82$ and spectral index $n_s = 0.97$.
[3]Energetics and synthesis method are the same as in Maraston (2005) and Maraston & Strömbäck (2011).

Here, we use an updated version of the Maraston et al. (2020) models, spanning a wider age range, with ages down to $\sim 3$ Myr, with a grid of 42 age values in total. The chemical composition of the models goes between $-2.25$ and 0.35 dex in [Z/H], allowing a [Z/H] grid of 9 values (see Hill et al. 2021). Population models are calculated for eight different values for the IMF slope below 0.6 M$_\odot$, ranging between 0.3 and 3.8 in the notation in which the Salpeter (1955)'s slope is 2.35, for each age and metallicity combination.

With the MaStar-based population models, we generate a synthetic spectrum for each stellar particle in the TNG50 galaxies [assuming the Kroupa (2002) IMF].

Stellar population models are a key input of galaxy formation simulations and 'forward-modelling' (Baugh 2006; Tonini et al. 2010; Gonzalez-Perez et al. 2014): they provide the link to the observables, and they are instrumental to obtain the physical properties of data, through spectral fitting. Consequently, the choice of the model is an essential part of the comparison between galaxy simulations and observed data. Our adoption of MaNGA-based population models ensures we use the same spectral properties in the simulations as well as in the interpretation of galaxy data. We are therefore able to exclude any bias that would be caused by the adoption of different spectral models. Moreover, as all spectra involved in our work have been obtained with exactly the same instrument and observational set-ups, we achieve the highest degree of consistency to start a meaningful comparison between data and simulations (Paper III).

### 2.3 The MaNGA galaxy survey

MaNGA (Bundy et al. 2015) is the largest Integral Field Spectroscopy (IFS) survey of galaxies to date. It observed 10 010 unique galaxies at a median redshift of $z \sim 0.03$ (Abdurro'uf et al. 2022) providing spatially resolved spectra for each of them. MaNGA is part of the SDSS-IV survey (Blanton et al. 2017) and concluded its observations in 2020 August.

The MaNGA IFS (Drory et al. 2015) was based around the SDSS 2.5-m telescope at Apache Point Observatory (Gunn et al. 2006) and utilizes the *SDSS*-BOSS spectrograph (Dawson et al. 2013; Smee et al. 2013), with a wavelength range from 3600 to 10 300 Å and an average spectral resolution $R \approx 1800$. In particular, the SDSS-BOSS spectrograph has a red and a blue camera, with a dichroic splitting the light around 6000 Å. In the blue channel, the resolution goes from $R = 1560$ at 3700 Å to $R = 2270$ at 6000 Å; in the red channel, the resolution goes from $R = 1850$ at 6000 Å to $R = 2650$ at 9000 Å. In Paper I, we explain how we mimic the resolution of the SDSS-BOSS spectrograph at different wavelengths when we generate both the synthetic spectra and the noise.

MaNGA has hexagonal-formatted fiber bundles, made from 2 arcsec-core-diameter fibers, conducting dithered observations with Integral Field Units (IFUs), which vary in diameter from 12″.5 (19 fibers) to 32″.5 (127 fibers) (see table 2 in Bundy et al. 2015). The hexagonal-formatted fiber bundles are mimicked in our forward modelling of the simulated galaxies, as explained in Paper I.

MaNGA is characterized by a spatial resolution of 1.8 kpc at the median redshift of 0.037 (Law et al. 2016). The MaNGA's characteristic fiber-convolved point-spread function (PSF) has a full width at half maximum (FWHM) of 2.5 arcsec (Law et al. 2015). For each MaNGA datacube, the 'reconstructed' PSF, or effective PSF (ePSF), is supplied in different bands (Law et al. 2016). We use the ePSFs in the different bands when generating the mock MaNGA-like galaxy datacubes (see Paper I).

The MaNGA galaxy sample is divided into a 'Primary' and a 'Secondary' sample, with a 2:1 split, for which the galaxy light sampling extends out to 1.5 effective radius ($R_{\rm eff}$) and 2.5$R_{\rm eff}$, respectively, (Wake et al. 2017). In this paper, we focus on building a Primary MaNGA-like sample from TNG50 simulated galaxies, see Section 4.

## 3 MOCK GALAXY INPUT AND CALCULATION

Here, we recapitulate our procedure to generate and analyse mock MaNGA galaxies, as introduced in Paper I.

### 3.1 Modelling the spectrum

Once a simulated galaxy is selected for post-processing, the first step is to model their stellar spectra. Our spectral modelling depends on the particle's age. If a stellar particle is younger than 4 Myr, we assume it to be a star-forming region and model its emission with the MappingsIII star-forming region models (MIII models, see Groves et al. 2008). For older ages, we use MaStar stellar population models (see Maraston et al. 2020, and Section 2.2). It is important to emphasize that the synthetic spectra are associated directly with the stellar particles in TNG50 by interpolating within the SSP model grid. This is different from the studies presented by Ibarra-Medel et al. (2018) and Sarmiento et al. (2023) in which the stellar particles in the simulations are assigned model spectra from the closest properties in the stellar population model template grid without interpolation, introducing a difference between 'intrinsic' and 'assigned' properties. This point will be reprised in Section 6.

We mimic an IFU observation to collect the light, generating a datacube with the MaNGA pixel size (0.5 arcsec) and a square FoV of 150 arcsec per side. Thanks to the use of the MaStar stellar population models, the synthetic datacube's spectral resolution and flux calibration are equal to MaNGA observations by construction. The virtual instrument is positioned along the $z$-axis of the cosmological volume in which the galaxy is identified. Note that observing the simulated galaxies with a line-of-sight (LOS) fixed to the cosmological $z$-axis effectively implies random viewing angles. An in-depth discussion of galaxy inclinations in iMaNGA sample will be done in Paper III.

### 3.2 Dust

Dust effects are included in the synthetic datacubes by reconstructing the attenuation curves spaxel by spaxel employing low-resolution radiative transfer simulations with SKIRT (Baes et al. 2011; Baes & Camps 2015). We define and discuss this original and fast way to exploit radiative transfer simulations in section 3.2.2 of Paper I. With SKIRT, we mimic an IFU observation with the same FoV and pixel size as for the synthetic datacubes, but with lower spectral resolution. It is important to underline that we do not assume any model for the attenuation curves: the attenuation curves are defined simply as the ratio between the signal in the spaxel with and without dust included in the radiative transfer simulations. The attenuation curves defined at lower spectral resolution are then interpolated on the wavelength array of the synthetic datacubes. The attenuation curves are then applied to the synthetic datacubes, spaxel by spaxel.

### 3.3 Kinematics

The kinematics are incorporated as follows. Spectra are Doppler-shifted and broadened according to stellar kinematics, from the TNG50 simulations (see section 3.2 in Paper I).

The iMaStar code, which performs these steps, can be found here: https://github.com/lonanni/iMaNGA.

### 3.4 Morphology

For the morphological analysis, we first obtain $r$-band SDSS-like images by applying the SDSS $r$-band filter and PSF to the synthetic datacubes. Images are then analysed with STATMORPH, a Sérsic 2D fitting code (Rodriguez-Gomez et al. 2019, see Paper I, section 3.3). The analysis with STATMORPH provides us with the effective radius $R_{\rm eff}$ which is needed to construct a mock MaNGA Primary sample.

### 3.5 Inclusion of observational effects

Once the $R_{\rm eff}$ values of the galaxies are known, we select the appropriate hexagonal fiber-bundle configuration that would be employed by MaNGA to collect the light from the galaxy within 1.5$R_{\rm eff}$. The available hexagonal diameters in the MaNGA set-up go from 12.5 to 32.5 arcsec, see Section 2.3. As discussed in Paper I, we do not simulate the detailed spatial sampling mechanics of MaNGA observations, as done in Bottrell & Hani (2022), for instance. Instead, we select all the spaxels within the MaNGA hexagonal-formatted fiber-bundle FoV [see also Nevin et al. (2021)] and mimic observational effects such as dithering and resulting covariances between spaxels by exploiting the reconstructed PSF (or effective PSF) provided for each of the observed MaNGA galaxies in different bands.

The effective PSF depends on the exposure time and the observing condition in the considered band, and includes also dithering effects (Law et al. 2016). The convolution of the datacubes with the effective PSF happens after the implementation of noise. The noise is modelled based on an analysis of the real wavelength-dependent SNR in MaNGA (see Paper I). Since the convolution happens after the inclusion of the noise in each spaxel, this results in the signal of adjacent spaxels being correlated, including also the noise. This indirect approach of mimicking the effects of the MaNGA IFS observations leads to significant savings in computing time. It is a simplification compared to reconstructing the detailed mechanics of the observations, as for instance done by Bottrell & Hani (2022). Both the reconstruction of the ePSF and the noise are based on MaNGA *LOGCUBE* output (Law et al. 2016). We refer the reader to section 3.4 of Paper I for more detail.

Following this approach, and thanks to the combination of TNG50 and MaStar Stellar Population models, we produce datacubes having the same spatial sampling, spatial resolution, spectral resolution, SNR as a function of the wavelength, flux calibration, and wavelength range of MaNGA observations.

### 3.6 Data analysis

We follow the procedure developed for the MaNGA Data Analysis Pipeline (Westfall et al. 2019) to analyse the iMaNGA galaxies. First, we employ the Voronoi algorithm of Cappellari & Copin (2003), with target S/N$_g$ > 10. We then run the penalized pixel-fitting algorithm (pPXF; Cappellari 2017), to reconstruct gas and stellar kinematics and model the emission lines.

We finally fit the mock spectra with FIREFLY (Wilkinson et al. 2017) to infer the stellar populations' properties, i.e. age, metallicity, mass, and age, and also reddening and the SFH. As we show in Paper I on two test galaxies with vastly different properties, our 'mocking' and fitting procedures do not alter the intrinsic properties of the TNG50 galaxies.

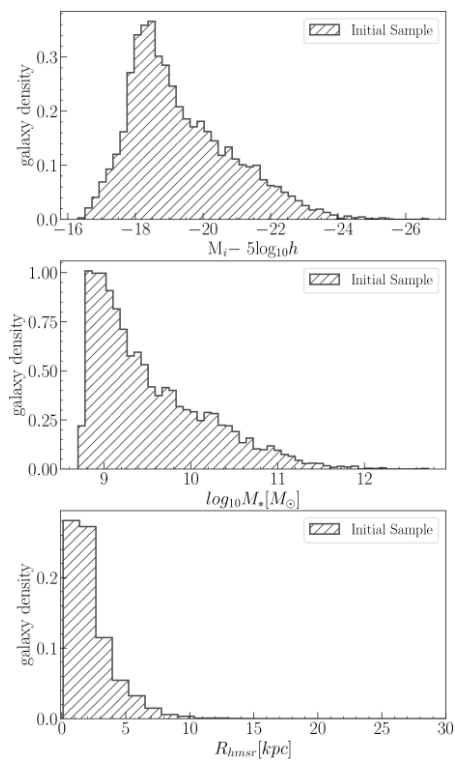

**Figure 1.** Distribution of (from top to bottom): *i*-band absolute magnitude (AB) (where $h = H_0/100\,\mathrm{km\,s^{-1}\,Mpc^{-1}}$ and $h = 1$), stellar mass and half-mass–stellar radius (as a proxy of a galaxy size), of the 'initial sample', i.e. the 48 248 TNG50 galaxies selected to lie in the MaNGA redshift range (i.e. 0.01–0.15) and to having more than 10 000 stellar particles.

## 4 MOCK CATALOGUE CONSTRUCTION

Here, we illustrate our method to construct the mock MaNGA catalogue from TNG50, i.e. our iMaNGA sample.

### 4.1 The initial sample of TNG50 galaxies

As explained in Section 2.1, the SUBFIND algorithm is run over all the saved snapshot of the Illustris and IllustrisTNG output cosmological volume. This algorithm identifies structures in the cosmological volumes as galaxies. To construct the iMaNGA sample from the TNG50 simulations, at first, we select all the TNG50 galaxies in the MaNGA redshift range (between $z \approx 0.15$ and $z \approx 0.01$, see Section 2.3). This corresponds to 10 snapshots, from snapshot 88 to snapshot 98. We remove all the galaxies with less than 10 000 stellar particles from this sample, to ensure that dust effects are sufficiently resolved (as done in Schulz et al. 2020). We also remove all the 'galaxies' which are flagged as spurious artefacts of the SUBFIND procedure (for more details see Genel et al. 2017; Pillepich et al. 2018b, and the Data Specification page for IllustrisTNG)[4].

We find that 48 248 galaxies in TNG50 satisfy these selection criteria. We shall refer to this sample of galaxies as the 'initial sample'. Fig. 1 displays the initial sample in (from top to bottom): *i*-band absolute (AB) magnitude, stellar mass, and half-mass–stellar radius $R_{\mathrm{hmsr}}$ as a proxy of galaxy sizes. We refer to the *i*-band AB magnitude as $M_i - 5\log_{10} h$, as in the NSA catalogue[5], which was used for the MaNGA target selection. As in the NSA catalogue, $h = H_0/100\,\mathrm{km\,s^{-1}\,Mpc^{-1}}$, and $h = 1$ throughout the paper. It can be appreciated how the majority of galaxies in the initial sample have a stellar mass around $9 \times 10^{10}\,\mathrm{M_\odot}$.

[4] https://www.tng-project.org/data/docs/specifications/
[5] http://nsatlas.org/data

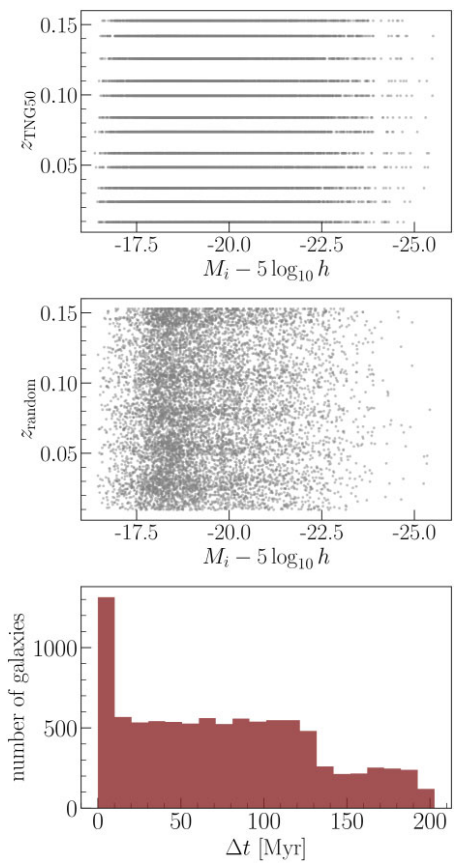

**Figure 2.** The initial sample of TNG50 galaxies in the magnitude–redshift plane, before and after randomizing their redshift (upper and central panel, respectively). The difference between these two redshift values, given as $\Delta t$, is shown in the bottom panel.

### 4.2 Obtaining a smooth spatial sampling

Since we want to recover a smooth distribution in spatial sampling, as in the MaNGA-Primary sample, we need to alter the discreteness of the TNG50 redshift sampling, due to the fact that – at low-$z$ – the TNG50 snapshots are output every 150 Myr, and make it a continuum sample. To this end, we associate to each galaxy in the initial sample (see Section 4.1) a new redshift, which we call $z_{\mathrm{random}}$. This redshift is randomly extracted from a uniform distribution with lower limit equal to the galaxy's snapshot redshift (we refer to it as $z_{\mathrm{TNG50}}$), and upper limit equal to the redshift of the previous, higher redshift snapshot. In other words, we allow a galaxy in a given snapshot to have a redshift between the redshift characterizing its snapshot and the redshift of the preceding snapshot. We do not change the redshift of galaxies at the upper redshift boundary, i.e. $z \approx 0.15$. In this way, we obtain a smooth distribution in spatial sampling, as for MaNGA datacubes. The results of the whole procedure are visualized in Fig. 2.

From now on, we consider the galaxies in the initial sample as characterized by $z_{\mathrm{random}}$, instead of their original redshift $z_{\mathrm{TNG50}}$. It should be noted that the new redshift $z_{\mathrm{random}}$ is solely used to construct a MaNGA-like catalogue, and to *observe* the galaxies in it when producing mock MaNGA-like observations (see Section 3). The ages of stellar particles as provided by TNG50 are not modified.

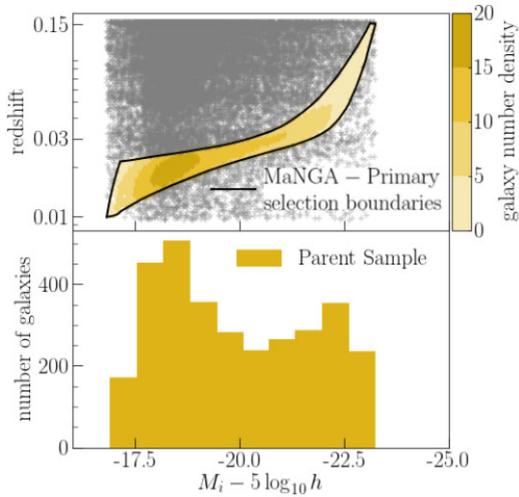

**Figure 3.** *Upper panel:* The distribution of the 3152 TNG50 galaxies in the parent sample in the magnitude–redshift space, colour coded by the galaxy number density, and the MaNGA-Primary sample selection boundaries (black solid line). The grey points represent the galaxies excluded from the sample because not in the MaNGA-Primary target. *Bottom panel:* the distribution in *i*-band magnitude of the parent sample.

### 4.3 Magnitude selection: the parent sample

The MaNGA sample selection is solely based on the galaxies' absolute *i*-band magnitude and redshift, with the final sample achieving an approximately flat distribution in the *i*-band magnitude (for more information see Yan et al. 2019).

The top panel of Fig. 3 presents the MaNGA-Primary sample selection boundaries (in black) (see Wake et al. 2017), and the TNG50 galaxies in the initial sample falling into it, identified by the new redshift $z_{\rm random}$. The bottom panel shows their distribution in the *i*-band magnitude. After imposing the Primary sample selection boundaries, we are left with 3152 TNG50 galaxies. We refer to this sample as the 'parent sample'.

Fig. 4 displays the magnitude, the stellar mass $M_*$, and the half-mass–stellar radius $R_{\rm hmsr}$ distributions (from top to bottom) for the initial (hatched-filled histograms) and the parent sample (yellow histograms). Excluding galaxies outside the MaNGA selection boundaries (grey points in Fig. 3) changes the galaxy distributions substantially. There is a significantly higher density of high-luminosity objects and high-mass objects once these selection criteria are applied. The size distribution, instead, remains largely unchanged.

### 4.4 The final iMaNGA sample

In order to achieve the final, flat distribution in *i*-band magnitude over the TNG50 parent sample (Section 4.3), we perform the following steps:

(i) We build the galaxy distributions in *i*-band magnitude and redshift of the parent sample (bottom panel of Fig. 3),

(ii) To each TNG50 galaxy in the parent sample, we associate a probability $p$ inversely proportional to their containing bin count, $N$ (bottom panel of Fig. 3);

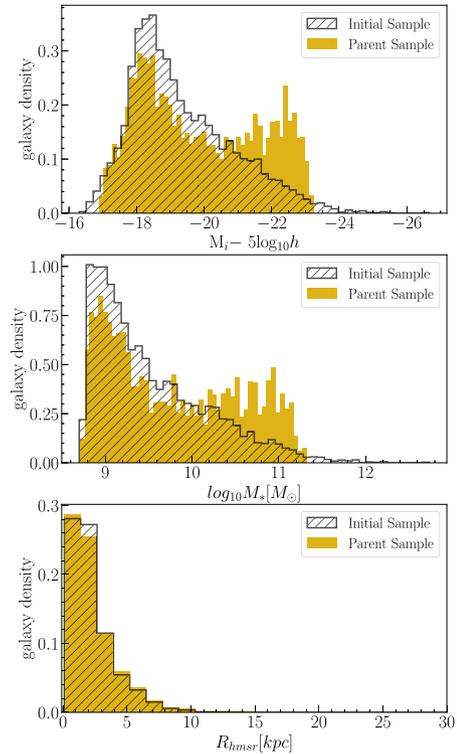

**Figure 4.** Comparison between the initial (hatch-filled histograms) and the parent (yellow histograms) samples of TNG50 galaxies, in *i*-band magnitude, stellar mass, and half-mass–stellar radius (from top to bottom panel).

(iii) We randomly extract unique galaxies from this sample, assuming a selection probability for each of them equal to $1/N$. In other words, the probability of being selected is larger in underpopulated bins.

The requirement to achieve a flat distribution constrains the number of galaxies, which ends up being about 1000 in our case. We will refer to this sample as the 'iMaNGA sample'. This is our final sample, which we then post-process and analyse following the method presented in Paper I and recapitulated in Section 3.

The top panel of Fig. 5 displays the MaNGA-Primary sample selection boundaries (in black), and the TNG50 galaxies in the iMaNGA sample, colour coded by number. The bottom panel shows the distribution in *i*-band magnitude in the initial (black hatch-filled histogram), parent (yellow histogram), and final iMaNGA (teal empty histogram) samples of TNG50 galaxies. Our method allows us to successfully generate a flat distribution in *i*-band magnitude. In Fig. 6, we compare these three samples in terms of stellar mass (top panel) and size (bottom panel, where the half-mass–stellar radius is considered a proxy of the galaxy size as usual), with the same meaning of symbols as in the bottom panel of Fig. 5. The iMaNGA sample of ∼1000 unique TNG50 galaxies is characterized by an approximately flat distribution in mass, as well as in *i*-band magnitude.

## 5 RESULTS

In this Section, we present the main characteristics of the iMaNGA sample, in particular the galaxy sizes, masses, and environmental densities (Section 5.1). Then, we report the outcome of the morphological analysis, based on mock *r*-band SDSS-like images (Section 5.2). Next, we present a direct comparison between some key characteristics of the iMaNGA and the Primary MaNGA samples (Section 5.3). Afterwards, we describe the results of the kinematic analysis over the entire sample, discussing our ability to recover the 'intrinsic' kinematics, as given by TNG50 (Section 5.4). At the end, we present the analysis of the stellar populations' properties and our ability to recover the 'intrinsic' stellar population properties from TNG50 (Section 5.5).

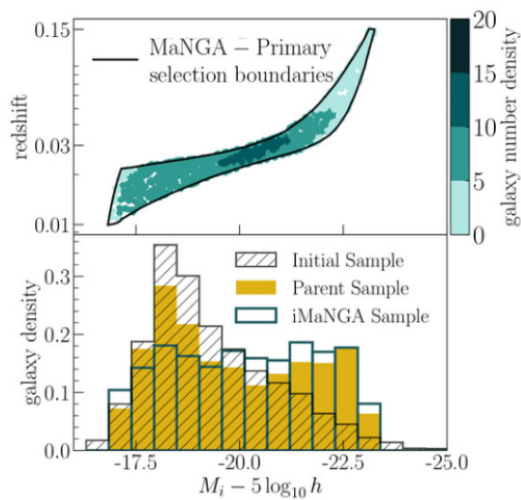

**Figure 5.** Redshift and magnitude distributions of the iMaNGA sample of TNG50 galaxies compared to the initial and parent samples. The solid black line defines the selection boundaries for the MaNGA Primary sample.

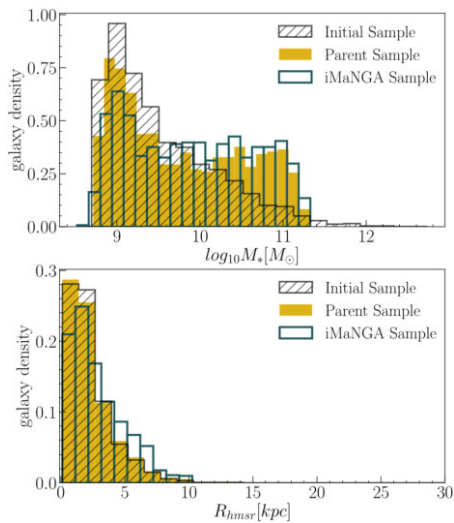

**Figure 6.** The distribution of TNG50 galaxies in the initial (hatch-filled black histograms), parent (yellow histograms), and iMaNGA (teal empty histograms) samples, in stellar mass and half-mass–stellar radius.

### 5.1 Characteristics of the iMaNGA sample

Fig. 7 shows the size (half-mass–stellar radius $R_{hsmr}$, top panel), the $i$-band magnitude ($M_i$, central panel), and the environmental density ($\log(1 + \delta)$, bottom panel) of the galaxies in the iMaNGA sample as a function of stellar mass ($M_*[10^{10} M_\odot]$) and colour coded by redshift.

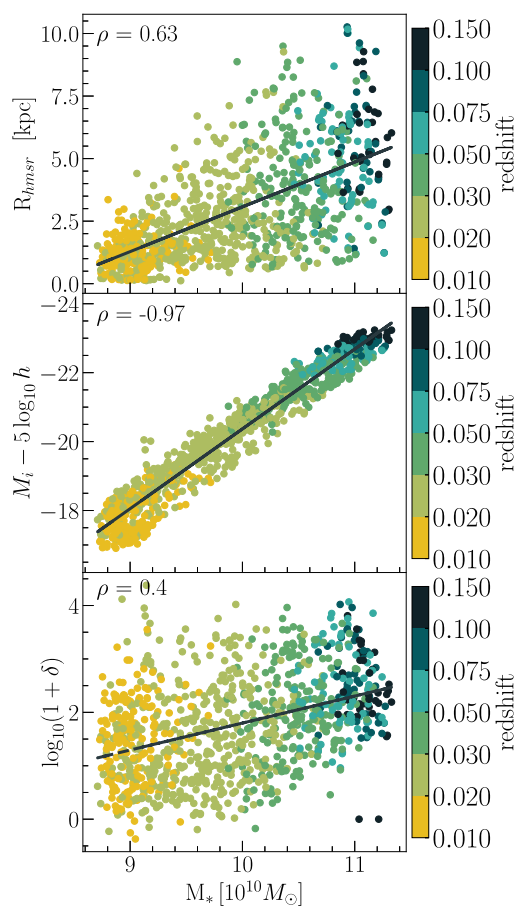

**Figure 7.** The distribution in half-mass–stellar radius, $i$-band magnitude, and environment, i.e. $\log(1 + \delta)$, (from top to bottom, respectively), as a function of stellar mass, colour coded by redshift, for the iMaNGA sample of TNG50 galaxies. The linear regression (black lines) and the Pearson correlation coefficient (in the upper left corners) are reported in each panel.

The environment is defined with the $N$th nearest neighbour method: the distance to the $N$th nearest neighbour, $d_N$, with N typically varying from 3 to 10, is used as a measure of the local galaxy (projected) overdensity (Muldrew et al. 2011; Etherington & Thomas 2015). The dimensionless overdensity, $1 + \delta$, is described by the following equation:

$$1 + \delta = 1 + \frac{\Sigma_N - <\Sigma>}{<\Sigma>}, \quad (1)$$

where $\Sigma_N$ is the surface number density described using the $N$th neighbour method ($\Sigma_N = N/\pi d_N^2$), while $<\Sigma>$ is the mean surface density of galaxies. We assume $N = 5$ and compute $<\Sigma>$ within the TNG50 snapshot the galaxies are in, projecting the subhaloes in each of the simulated volumes along the $z$-axis of the cube. As a proxy of the galaxy environment, we use $\log(1 + \delta)$ (for more details see Paper I). In each panel of Fig. 7, we report the linear regression (black line) and the Pearson correlation coefficient $\rho$.

As expected, galaxy mass and $i$-band magnitude are closely correlated. However, there are also significant correlations between stellar mass and the half-mass radius as well as the environmental density. These correlations are consistent with observations of real galaxies (see Kauffmann et al. 2004; Bernardi et al. 2010; Li et al. 2018). As expected, due to the design of the MaNGA target selection (see Yan et al. 2019), there is a strong redshift bias, with more massive galaxies lying at higher redshift in order to fit into the IFU apertures. Our iMaNGA sample reproduces this effect.

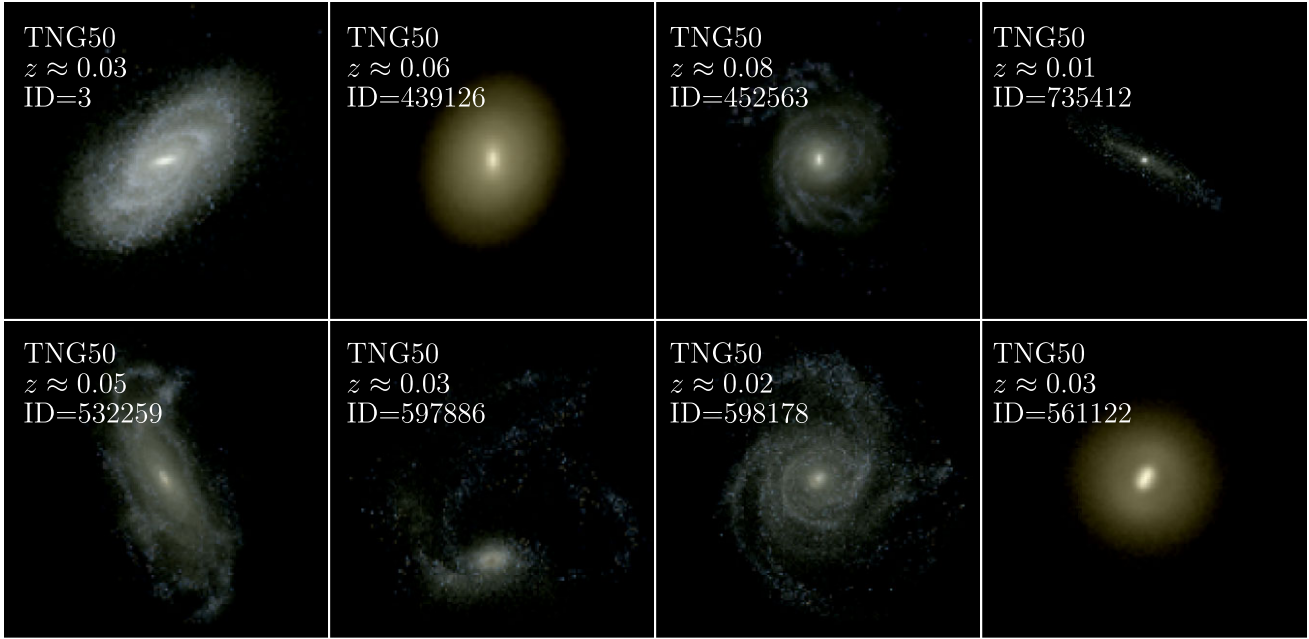

**Figure 8.** RGB images of 8 galaxies in the iMaNGA sample (see Section 4.4). The RGB images are constructed from their synthetic datacubes, generated with the method presented in section 3.2 in Paper I (see Section 3 here). Different angular sizes and morphologies can be appreciated. The redshift and the ID of the galaxies in TNG50 are reported in the upper-left corner for each galaxy. Their key properties are stated in Table 1.

**Table 1.** Selected properties of the 8 galaxies in the iMaNGA sample. From left, the redshift, the magnitude in the $i$ band, the stellar mass, and the $R_{\rm HMSR}$ as given by the TNG50 data, while the environment $(1 + \delta)$, the effective radius $R_{\rm eff}$ and the Sérsic index $n$ are computed as described in Paper I, from $r$ band SDSS mock images.

| snap-id | $z_{\rm random}$ | $M_i$ | $\log_{10} M_*$ ($M_\odot$) | $R_{\rm HMSR}$ (kpc) | $(1 + \delta)$ | $r$-band $R_{\rm eff}$ (kpc) | $r$-band $n$ |
|---|---|---|---|---|---|---|---|
| 96–3 | 0.034 | −22.13 | 10.55 | 3.3 | 1.94 | 4.86 | 0.94 |
| 94–439126 | 0.072 | −22.35 | 11.11 | 3.22 | 1.81 | 4.0 | 2.39 |
| 92–452563 | 0.093 | −22.85 | 10.95 | 6.11 | 2.7 | 8.15 | 1.48 |
| 98–735412 | 0.023 | −18.53 | 9.11 | 2.67 | 0.39 | 2.39 | 0.49 |
| 95–532259 | 0.05 | −21.88 | 10.53 | 7.62 | 3.41 | 13.66 | 0.62 |
| 97–597886 | 0.033 | −20.76 | 9.95 | 8.87 | 2.84 | 8.64 | 0.32 |
| 97–598178 | 0.034 | −21.35 | 10.31 | 4.39 | 2.65 | 4.75 | 0.58 |
| 96–561122 | 0.046 | −21.98 | 10.97 | 1.78 | 1.89 | 1.26 | 2.16 |

### 5.2 Morphological analysis

As said earlier (and explained in Paper I), morphologies are obtained from the $r$-band images. Fig. 8 displays the RGB images of 8 galaxies in our iMaNGA sample, to showcase the variety of morphologies reproduced by the TNG50 simulations, and therefore also present in the iMaNGA sample. The RGB images are obtained from the synthetic datacubes, with a square FoV of 150 arcmin as side length, including the effects of dust. Note that we do not include any observational effects, such as noise or PSF, as these images do not have a scientific purpose, being solely meant to visualize the variety of morphology and inclination in iMaNGA.

Table 1 reports the key properties of these galaxies: snapshot id, redshift $z_{\rm random}$, $i$-band magnitude, stellar mass, HMSR, environment (see Section 5.1), effective radius, and Sérsic index. Comparing the Sérsic indices with Fig. 8, we notice how galaxies showing spiral structures (ID 3, ID 452563, ID 532259, and ID 598178) have $n \sim 1$, while elliptical-looking galaxies (ID 439126, ID 561122) have $n > 2$.

Fig. 9 shows the distribution of Sérsic index ($n$) in iMaNGA. The top panel presents the entire catalogue (teal histogram). In the central panel, the catalogue is split into galaxies with and without dust particles (grey and black dashed histogram). In the bottom panel, the catalogue is divided into galaxies with and without star-forming particles (grey and black dashed histogram). The distribution in Sérsic index of the galaxies with star-forming and dust particles peaks at low $n$ values ($n \sim 1$). Indeed, it is expected to have star formation and dust in disc-like and irregular galaxies (Lianou et al. 2019).

We note that the overall distribution of the Sérsic index in iMaNGA has a peak around $n \sim 1$. A similar peak is also present in the distributions of the Sérsic index in the MaNGA catalogue. However, the Sérsic index distribution in MaNGA is bimodal, with another, slightly smaller peak around 4 (see fig. 12 in Fischer, Domínguez Sánchez & Bernardi 2018). This second peak is also present in the iMaNGA sample, but significantly smaller. This is the direct consequence of a paucity of elliptical galaxies in the underlying TNG50 catalogue. We will discuss this point further in the following section.

### 5.3 Comparison with the MaNGA-Primary sample

In this section, we present a comparison between the main characteristics of the MaNGA-Primary sample and the iMaNGA catalogue. All the properties of the MaNGA-Primary sample, i.e. total stellar mass, effective radius, Sérsic index, and redshift, are retrieved from DRPALL MaNGA data (Law et al. 2016). The total stellar mass of a galaxy in the iMaNGA sample is intrinsic, hence is directly computed as the sum of the stellar particles used to construct the synthetic datacube. Other quantities like Sérsic index and effective radius, instead, are derived from the synthetic iMaNGA images.

#### 5.3.1 Angular size and spatial resolution

In Fig. 10, we present the distributions of angular size (left-hand panels), spatial resolution expressed in kpc (central) and in terms of the effective radius (right-hand panels) for MaNGA (top panels) and iMaNGA (bottom panels). The figure mimics fig. 5 of Wake et al. (2017), including the fine binning in the stellar mass range $8.9 \leq \log_{10} M_*/M_\odot \leq 11.3$. The vertical lines illustrate the radius of the hexagonal FoV of the smallest and biggest fibre-bundle configurations used in MaNGA observations.

A slight shift towards smaller angular sizes in iMaNGA with respect to MaNGA is noticeable in the left-hand panels, which is most pronounced for low-mass galaxies. The tails to large angular sizes are very similar between simulated and real sample, though. This plot demonstrates the necessity of the redshift cut in both the MaNGA and the iMaNGA samples to ensure coverage of $1.5R_{\rm eff}$ by the 5 hexagonal FoVs in MaNGA.

The central panels show spatial resolution, calculated from the nominal angular resolution element (2.5 arcsec, both in MaNGA and iMaNGA) and galaxy redshift. The distributions for MaNGA and iMaNGA are very similar. Both peak at around 1 kpc, and exhibit a tail to significantly lower spatial resolution of up to 6 kpc in the most massive galaxies.

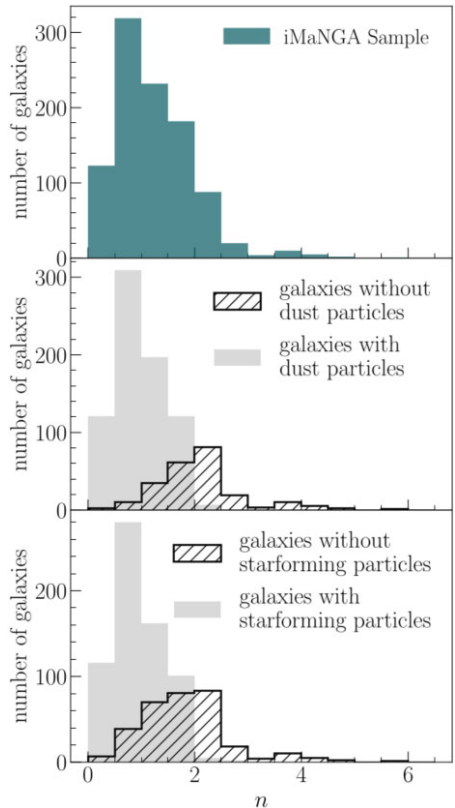

**Figure 9.** The distribution of the galaxies' Sérsic index $n$ as computed by running STATMORPH over $r$-band SDSS-mock images (see Section 3) for all the galaxies in the iMaNGA sample. *Upper panel*: the $r$-band Sérsic index distribution for all the galaxies in the iMaNGA sample (see Section 4.4). *Central panel*: the $r$-band Sérsic index distribution for the sample, split into galaxies where dust particles are present (grey histogram) and where are not (black hatch-filled histogram). *Bottom panel*: the $r$-band Sérsic index distribution, dividing the sample into galaxies with star-forming particles (grey histogram) and without (black hatch-filled histogram).

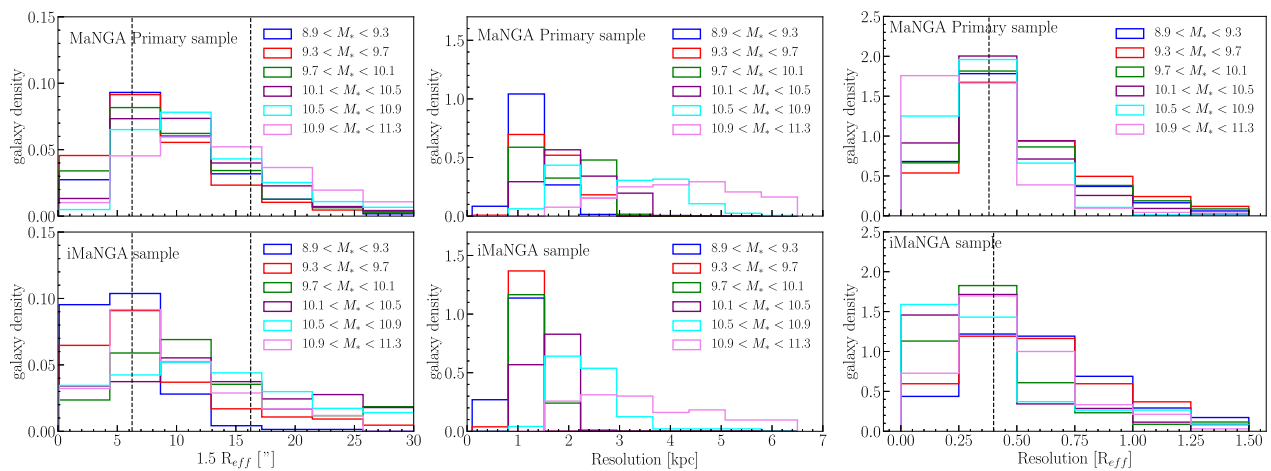

**Figure 10.** Comparison of angular size (left-hand panels), resolution in terms of kpc (central panels) and in terms of the effective radius (right-hand panels), of the iMaNGA sample, generated from TNG50 as described in this paper, to the MaNGA-Primary sample. This plot reproduces the plots in Wake et al. (2017). The information for the MaNGA-Primary sample galaxies is retrieved from DRPALL MaNGA data (Law et al. 2016), as for all the other plots in this Section. The vertical lines report the smallest and biggest FoV of the 5 hexagonal fiber-bundles configurations used in MaNGA observations (see Section 2.3).

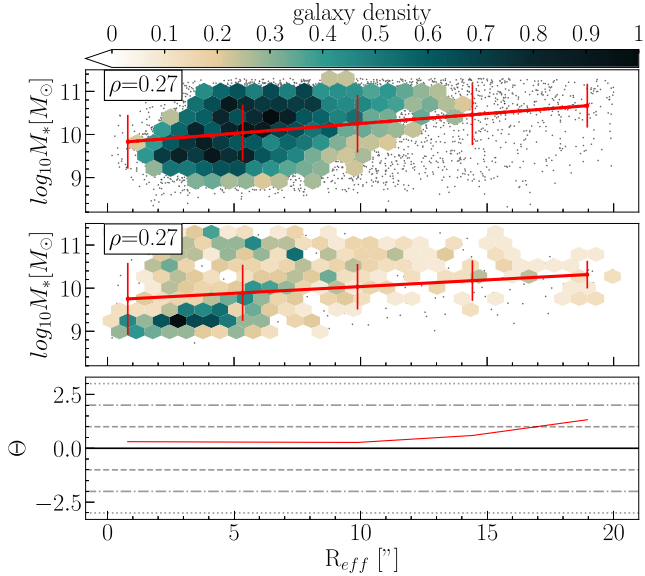

**Figure 11.** Stellar mass as a function of effective radius in the MaNGA-Primary (upper panel) and iMaNGA (central panel) samples. For both samples we show the median in five different equally sized bins in effective radius (red diamonds), the linear regression (red dotted line), the standard deviation (red error-bars), and the Pearson coefficient $\rho$. Colours represent the galaxy density in the tassels and the small black points show all galaxies in the samples. *Bottom panel*: normalized residuals (red line), and the $1\sigma$, $2\sigma$, and $3\sigma$ level intervals (dashed grey lines).

The right-hand panels show the distribution of spatial resolution in terms of $R_{\rm eff}$. As for the MaNGA-Primary sample, these distributions in the iMaNGA sample are largely independent of galaxy mass. Also, the median spatial resolutions are very similar: $0.37 R_{\rm eff}$ for the MaNGA-Primary sample, and $0.40 R_{\rm eff}$ for the iMaNGA sample.

We conclude that, overall, the iMaNGA sample reproduces the trends of the MaNGA-Primary sample for both angular size and spatial resolution well.

### 5.3.2 Stellar mass and effective radius

Fig. 11 shows a density plot of total stellar mass as a function of the effective radius for both the MaNGA-Primary sample (top panel) and the iMaNGA sample (central panel). As in Fig. 10, we select galaxies with a stellar mass in the range $8.9 \leq \log_{10} M_*/M_\odot \leq 11.3$. Coloured hexagons represent galaxy densities, and the small black points show individual galaxies. The red diamonds represent the median of the total stellar mass along 5 equally sized bins in effective radius, the red error-bars represent the standard deviation in each bin, and the red dashed line shows the linear regression with the Pearson correlation coefficient in the top-left corner. The bottom panel shows the residual calculated as

$$\Theta = \frac{O_{\rm MaNGA} - O_{\rm iMaNGA}}{\sqrt{\sigma_{\rm MaNGA}^2 + \sigma_{\rm iMaNGA}^2}}, \quad (2)$$

where $O_{\rm MaNGA}$ and $O_{\rm iMaNGA}$ represent the median of total stellar mass along the 5 equally sized bins considered for the two samples, and $\sigma_{\rm MaNGA}$ and $\sigma_{\rm iMaNGA}$ represent the standard deviation in each of the bins for the two samples. The correlation between effective radius and stellar mass is well recovered in the iMaNGA sample, in particular at smaller angular size.

### 5.3.3 Sérsic index

Fig. 12 presents a density plot of the Sérsic index as a function of effective radius (left-hand panels) and stellar mass (right-hand panels), both for the MaNGA-Primary sample (top panels) and the iMaNGA sample (central panels). We show all galaxies in the mass range $8.9 \leq \log M_*/M_\odot \leq 11.3$. The red diamonds represent the median along 5 equally sized bins with the red error-bars showing the median standard deviation within each bin. We also display the linear regression for the median values (red dashed lines), the Pearson correlation coefficient (in the upper left corners). With black points, we represent all the galaxies in the catalogues. In the bottom panels, we report the normalized residual between the two samples, computed as from equation (2) where $O_{\rm MaNGA}$ and $O_{\rm iMaNGA}$ represent the median of the Sérsic index, and $\sigma_{\rm MaNGA}$ and $\sigma_{\rm iMaNGA}$ represent the standard deviation in each of the bins.

The MaNGA-Primary sample shows a trend of Sérsic index slightly increasing with effective radius, while, the paucity of galaxies with high Sérsic indices leads to the opposite trend in the iMaNGA sample. The right-hand panels show that a correlation exists also between Sérsic index and total stellar mass for both the MaNGA and the iMaNGA sample. The correlation is weaker in iMaNGA, though.

Our results are in agreement with the study by Huertas-Company et al. (2019) who compare mock SDSS $r$-band images from TNG100 to SDSS. In particular they find that, although the observed mass–size relation is well recovered by TNG100, both in normalization and in slope, SDSS is dominated by elliptical galaxies at the high-mass end while TNG100 is dominated by late-type systems, with a lack of lenticular galaxies at intermediate masses and ellipticals at the high-mass end. This discrepancy is not due to volume, as shown by re-sampling the SDSS results over smaller volumes.

Similar results are also presented in Rodriguez-Gomez et al. (2019).

### 5.4 Kinematic analysis

The kinematic analysis of iMaNGA follows the procedure of the MaNGA DAP. In brief (see Paper I for details):

(i) We apply the Voronoi binning scheme to meet a given SNR threshold (i.e. a minimum SNR of about 10 in the $g$-band images), averaging over neighbouring spaxels. In this way, we define the Voronoi tessellation (Cappellari & Copin 2003) in each iMaNGA datacube. This step is necessary, both for real and mock MaNGA-like datacubes, since accurate measurements of stellar kinematics and stellar population characteristics need a good SNR to be extracted without bias. Where a good SNR cannot be reached, a mask is applied to not analyse those spaxels any further;

(ii) We run PPXF (Cappellari 2017) to obtain the stellar kinematics, as well as the gas kinematics and the emission lines best fit (see Belfiore et al. 2019; Westfall et al. 2019). The MILES-HC libraries are used as templates by PPXF for the stellar continuum, while the templates for the emission lines are constructed as explained in Westfall et al. (2019), Section 9.

In total we analyse ~8 million Voronoi tassels in the iMaNGA sample.

Fig. 13 reports examples of a fit with PPXF to the stellar continuum

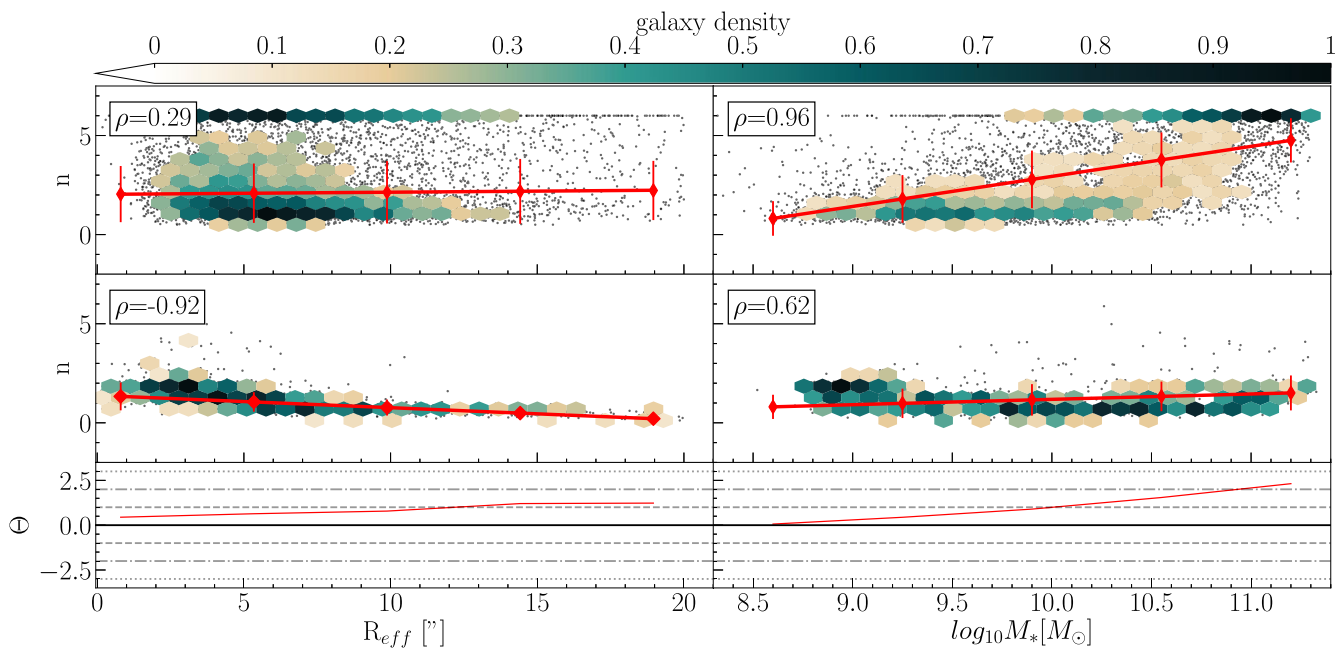

**Figure 12.** As in Fig. 11, for the distribution of the galaxies' Sérsic index $n$ as a function of the $r$-band $R_{\rm eff}$ and the total stellar mass, both for the MaNGA-Primary sample (*upper panels*) and the iMaNGA sample (*central panels*). *Bottom panels*: normalized residuals (red line), and the $1\sigma$, $2\sigma$, and $3\sigma$ level intervals (dashed grey lines). Symbols, curves, and colours as in Fig. 11.

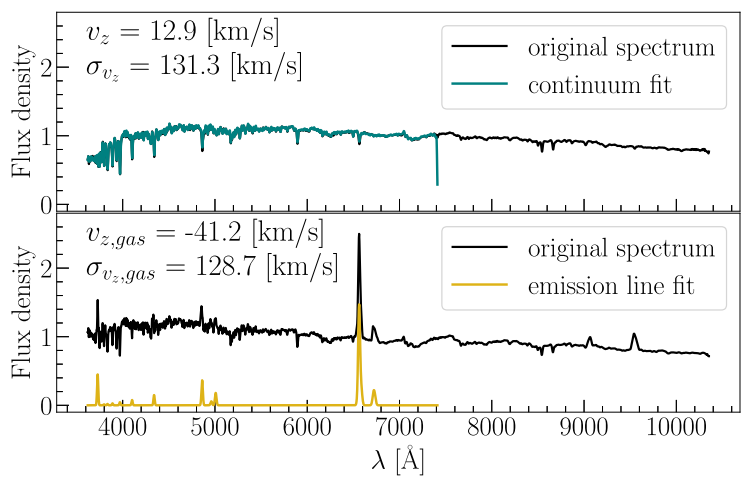

**Figure 13.** Examples of fit with PPXF for two spectra in the iMaNGA sample. *Upper panel*: the original spectrum (black line) and the best-fitting spectrum (teal line) determined by PPXF for the stellar continuum. In the upper left corner, the best-fitting values for the stellar velocity ($v_z$) and velocity dispersion ($\sigma_{v_z}$) are stated. *Bottom panel*: the original spectrum (black line) and the best-fitting spectrum (golden line) determined by PPXF for the emission lines.

(upper panel) and to the gas component (bottom panel); the best-fitting values for the stellar and gas kinematics are labelled in the figure. Fig. 14 displays the stellar peculiar velocity maps for the 8 galaxies in the iMaNGA sample, presented in Fig. 8 and Table 1, as recovered by running PPXF over the stellar continuum in each Voronoi tassel. In Fig. 15, we show the stellar velocity dispersion maps for the same galaxies, as recovered from the PPXF analysis. A variety of kinematical states, FoVs, Voronoi binning tessellation and masking can be appreciated. Comparing these with Fig. 8 and the Sérsic indices listed in Table 1, we notice how elliptical galaxies (ID 439 126 at $z \sim 0.06$, ID 561 122 at $z \sim 0.03$) have $n > 2$ and the highest values of stellar velocity dispersion, while still presenting some pattern of rotation. Disc-like galaxies (ID 3 at $z \sim 0.03$, ID

452 563 at $z \sim 0.08$, ID 532 259 at $z \sim 0.05$, ID 598 178 at $z \sim 0.02$), instead, have $n \sim 1$, exhibit a clear rotational velocity pattern, and a lower velocity dispersion compared to the ETGs.

In Paper I, we presented an initial comparison between the 'intrinsic' kinematics (peculiar stellar velocity and stellar velocity dispersion), i.e. determined directly from the simulations calculated from equation (5) in Paper I, and the 'recovered' kinematics obtained from running PPXF over the Voronoi-binned mock datacubes, for two example galaxies. We found that residuals are compatible with zero at the 68 per cent confidence level, with no systematic bias for the two iMaNGA datacubes.

We now repeat this comparison, this time for all the tassels in the iMaNGA sample (∼8 million) to test our ability to recover the input values statistically. Fig. 16 displays the distributions of the residuals (red histograms), both for the stellar velocity (i.e. $\Delta v_z = v_{z,\text{pPXF}} - v_{z,\text{TNG50}}$) and the stellar velocity dispersion (i.e. $\Delta \sigma_{v_z} = \sigma_{v_z,\text{pPXF}} - \sigma_{v_z,\text{TNG50}}$) over all the tassels in the iMaNGA sample. We also report the 0.16, 0.5, and 0.84 quartiles (vertical black lines). The quartiles for the $\Delta v_z$ distribution are equal to $q_{16} = -11.24$, $q_{50} =$ $-2.61$, and $q_{84} = 6.35$; for the $\Delta \sigma_{v_z}$ are $q_{16} = -4.04$, $q_{50} = 6.16$, and $q_{84} = 16.94$.

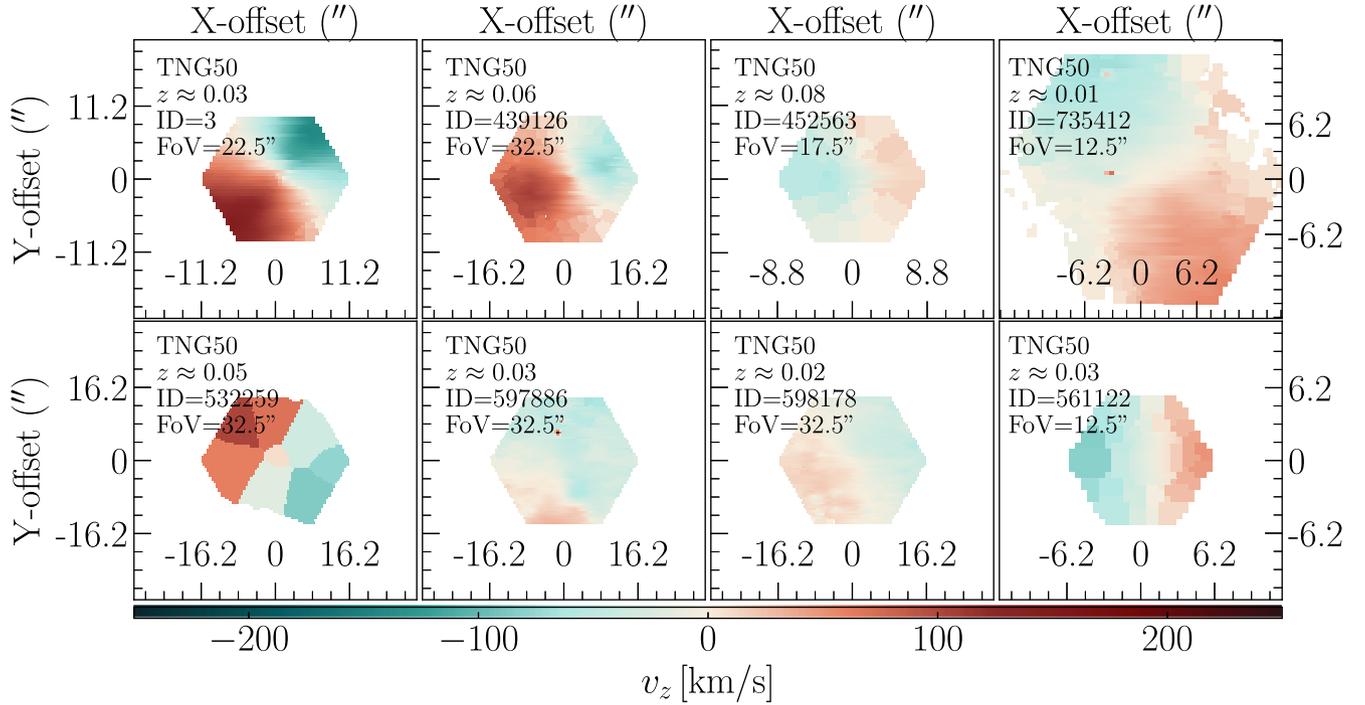

**Figure 14.** Stellar velocity maps, obtained with PPXF, for the 8 galaxies in the iMaNGA sample from Fig. 8 and Table 1. Maps are colour coded by the stellar peculiar velocity along the LOS. Galaxy redshifts and IDs are stated in the plots, as well as the diameter of the hexagonal MaNGA FoV the galaxy falls in, which is used to 'observe' the galaxy, see Section 3.

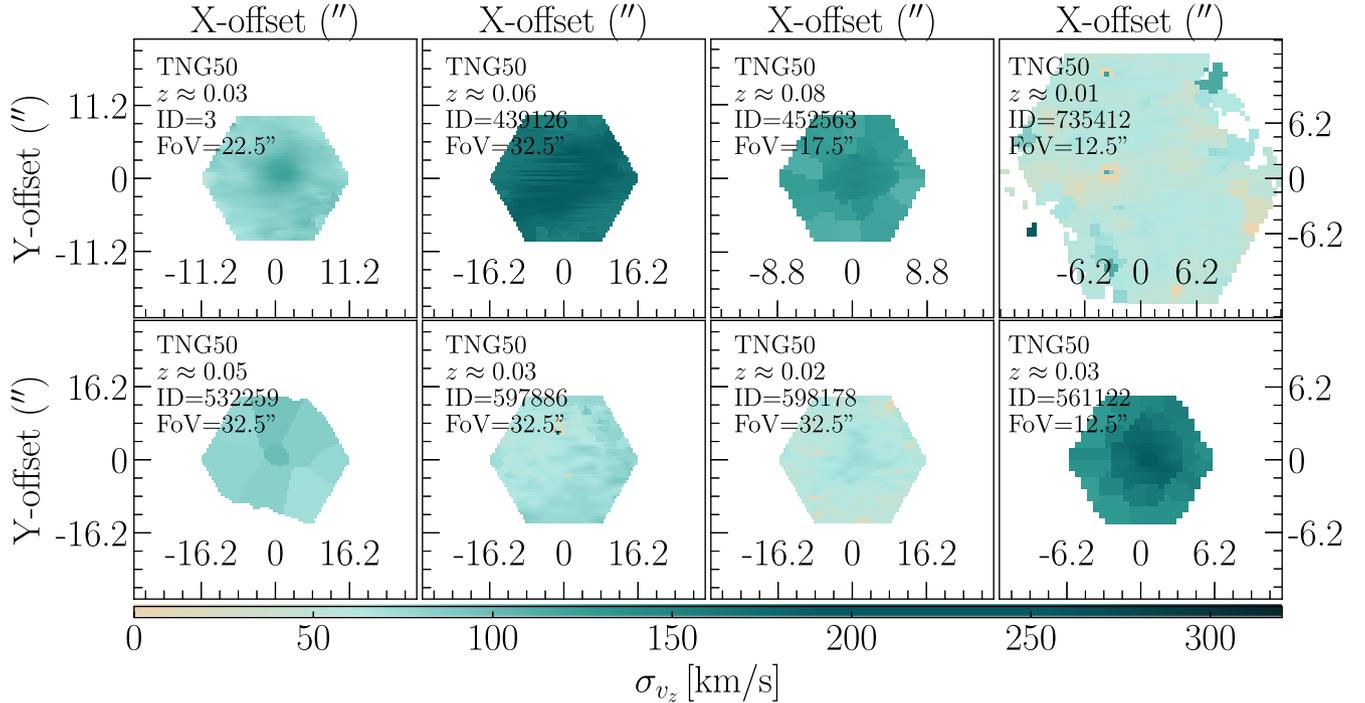

**Figure 15.** As in Fig. 14 for the stellar velocity dispersion.

At the $1\sigma$ level, the residuals are compatible with zero, hence we recover unbiased measurements of the stellar velocity and the stellar velocity dispersion along the LOS.

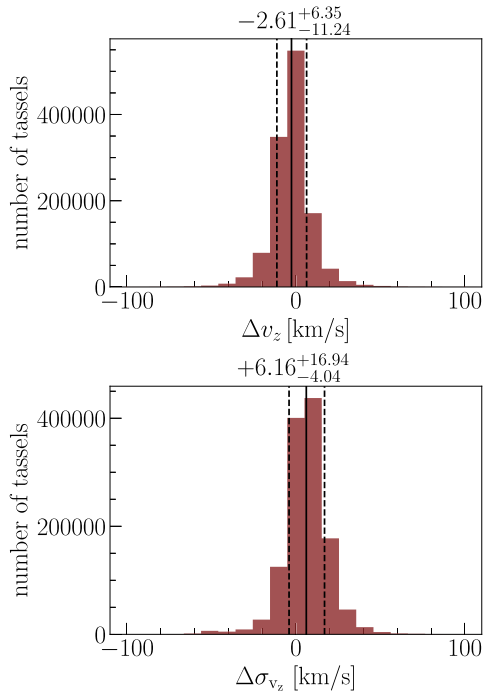

**Figure 16.** The distributions of residuals (red histograms), both for the stellar velocity ($\Delta v_z = v_{z,\text{pPXF}} - v_{z,\text{TNG50}}$) and stellar velocity dispersion ($\Delta \sigma_{v_z} = \sigma_{v_z,\text{pPXF}} - \sigma_{v_z,\text{TNG50}}$) over all the tassels in iMaNGA. We report the 0.16, 0.5, and 0.84 quartiles of the residual distributions (vertical black lines). Residuals are compatible with zero over all tassels at the 68 per cent confidence intervals.

Fig. 17 presents recovered stellar velocity dispersion as a function of intrinsic stellar mass (green circles). The median in 5 equally sized bins in stellar mass (black diamonds) and the standard deviation in each bin (black error bars) are also shown. The black line is the linear regression line for the median values, and the Pearson correlation coefficient is given in the top-left corner. A robust linear correlation between the stellar mass and $log_{10}\sigma_{v_z}$ is found, in good agreement with both real observations (e.g. Zahid et al. 2016) and other simulations (e.g. Pillepich et al. 2019).

In closing this section, we would like to emphasize that our analysis naturally considers random orientation, since we observe the galaxies with a fixed LOS along the $z$-axis of the cosmological volume, and the simulated galaxies are randomly orientated in it. Different orientations can be appreciated in Fig. 8.

### 5.5 Stellar population analysis

As discussed in Paper I, after following the steps in the DAP, we can proceed to perform full spectral fitting of population models to the iMaNGA galaxies' datacubes in order to derive stellar population properties. The Voronoi tassels, stellar kinematics, and emission-line best-fits are used as input to the full spectral fitting code FIREFLY (Wilkinson et al. 2017), using the same strategy detailed in Neumann et al. (2021, 2022) for the analysis of real MaNGA galaxies adopting MaStar stellar population models as fitting templates. The fitting results for the MaNGA galaxies are publicly available as value-added-catalogue (Neumann et al. 2021)[6].

---
[6] https://www.sdss4.org/dr16/manga/manga-data/manga-firefly-value-added-catalog/

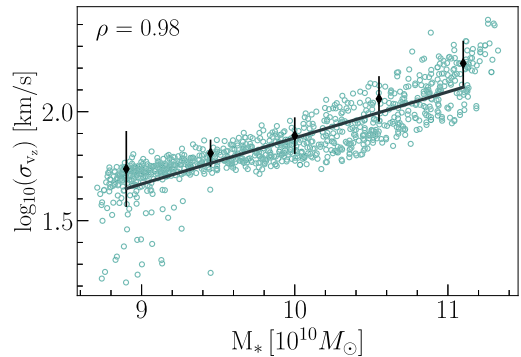

**Figure 17.** Average stellar velocity dispersion along the LOS as a function of stellar mass in the iMaNGA sample (green circles). Also shown is the median of the average stellar velocity dispersion in 5 equally sized bins in stellar mass (black diamonds) and the median standard deviation in each bin (black errorbars). The linear regression (black solid lines) and the Pearson correlation coefficient $\rho$ are reported.

Based on the full-spectral fitting with FIREFLY, fundamental galaxy properties are recovered including mass-weighted and light-weighted metallicities and ages. These properties are essential to investigate the formation and evolution of galaxies. Fig. 18 exemplifies a fit with FIREFLY to the stellar continuum for a relatively young stellar population (top panel) and an old stellar population (bottom panels). The best-fitting values for the stellar age and metallicity ([Z/H]) are reported in the top-left corners of the figure.

#### 5.5.1 Mass-weighted metallicity and age maps

Running FIREFLY over the mock IFU datacubes, we are able to spatially resolve the stellar population properties. Therefore, as in e.g. Neumann et al. (2022), we are in the position to produce maps of stellar properties for the iMaNGA datacubes.

Figs 19 and 20 show maps of recovered stellar metallicity and age for the eight galaxies presented in Fig. 8. Here, we report the mass-weighted metallicity and age, i.e. [Z/H]$_{\text{MW}}$ and $\log_{10}(\text{age})_{\text{MW}}$ (Gyr). We notice that elliptical galaxies (ID 439 126 at $z \sim 0.06$, ID 561 122 at $z \sim 0.03$) are metal-rich and old overall as expected. The iMaNGA disc galaxies (ID 3 at $z \sim 0.03$, ID 452 563 at $z \sim 0.08$, ID 532 259 at $z \sim 0.05$, ID 598 178 at $z \sim 0.02$), instead, have generally young stellar populations and show a metal-rich stellar component in the centre, with a steep drop in metallicity towards larger radii.

In Paper I we compared intrinsic age and chemical composition, calculated directly from the stellar particle data in TNG50, with the values recovered from FIREFLY for two iMaNGA galaxies, finding that residuals were consistent with zero at the 68 per cent confidence level with no systematic bias. Here, we repeat this comparison for the entire iMaNGA catalogue, thereby testing if, after the post-processing of the TNG50 galaxies, the 'intrinsic' information is correctly retrieved with our analysis.

To this end, weighted 'intrinsic' properties for the same Voronoi grid are constructed directly from the simulated stellar particles. The weighted properties in each tassel are:

$$\theta_{\text{W,tassel}} = \frac{\sum_{i=1}^{N_*} W_{*,i} \times \theta_i}{\sum_{i=1}^{N_*} W_{*,i}}, \quad (3)$$

where $\theta_{\text{W,tassel}}$ is either the mass-weighted metallicity ([Z/H]$_{\text{MW}}$), the light-weighted metallicity ([Z/H]$_{\text{LW}}$), the mass-weighted age ($Age_{\rm MW}$), or the light-weighted age ($Age_{\rm MW}$) in the selected tassel; $\theta_i$ is the age or metallicity of the $i$-particle of the considered tassel with weight, either mass or light, $W_{*,i}$; $N_*$ is the number of stellar particles in the selected tassel. For the mass-weight, i.e. $M_{*,i}$, we simply consider the stellar particle mass as provided by TNG50; for the light-weight, i.e. $L_{*,i}$, we consider the total luminosity of the stellar spectrum, either from MaStar or MIII, depending on the stellar particle's age.

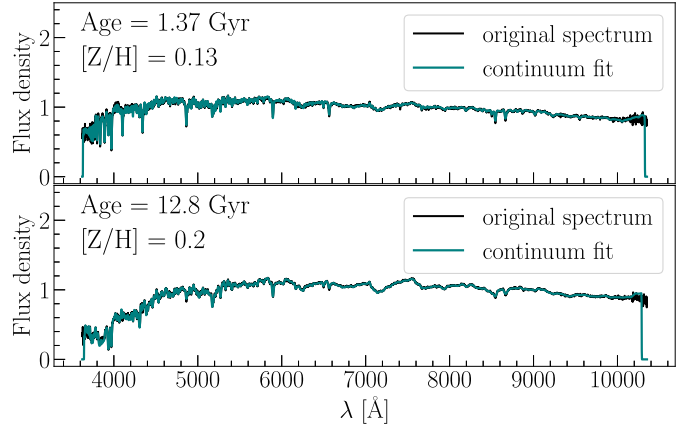

**Figure 18.** Examples of spectral fits for iMaNGA galaxies obtained with FIREFLY adopting the MaStar stellar population models as fitting templates. Original and best-fitting spectra determined by FIREFLY (black and teal lines, respectively) are shown for a relatively young stellar population (upper panel) and an older one (bottom panel). In the upper left corners, the best-fitting values for the stellar age and stellar metallicity ([$Z/H$]) are stated.

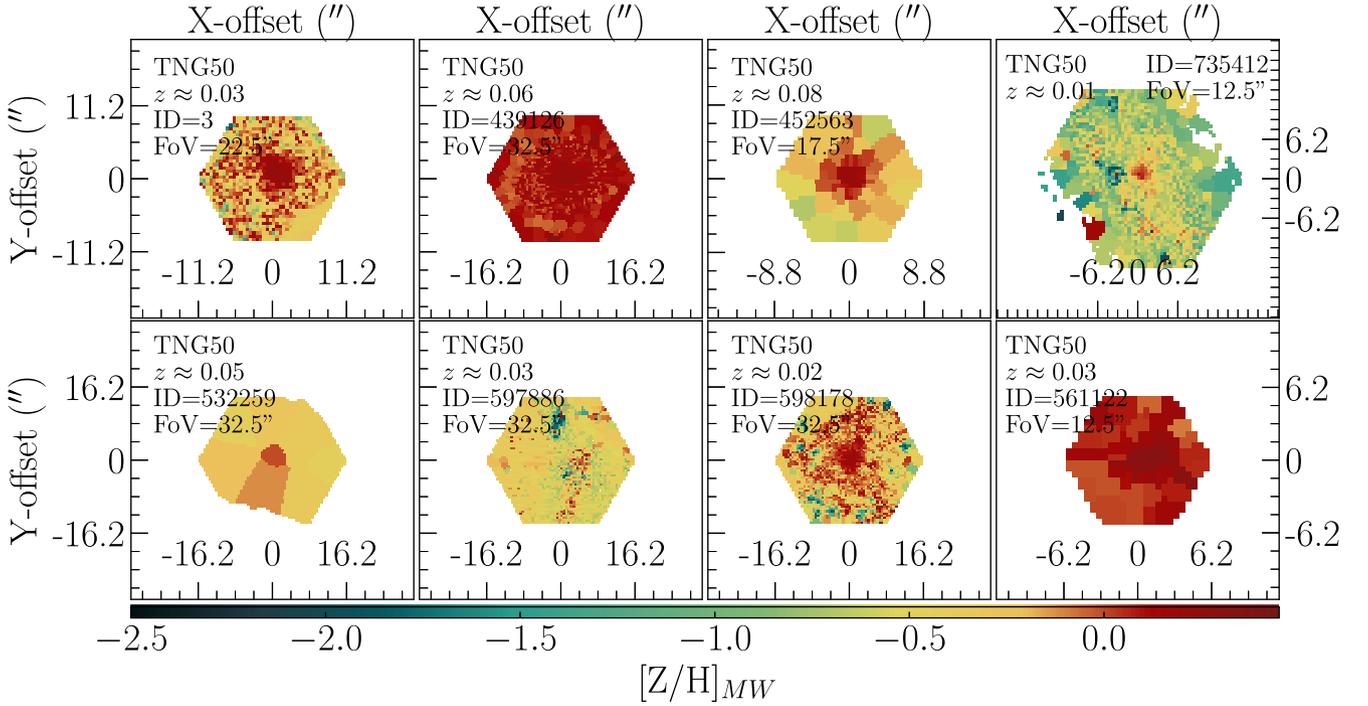

**Figure 19.** Stellar metallicity maps, measured by FIREFLY, for 8 galaxies in the iMaNGA sample, presented in Fig. 8. The redshift and the ID of the galaxies in TNG50 are listed in the upper-left corner for each galaxy, as well as the MaNGA FoV the galaxy falls in.

Fig. 21 reports the residual distributions for the metallicity (i.e. $\Delta[Z/H] = [Z/H]_{\rm Firefly} - [Z/H]_{\rm TNG50}$, top panel), and the residual distributions for the age ($\Delta \log_{10}(Age) = \log_{10}(Age)_{\rm Firefly} - \log_{10}(Age)_{\rm TNG50}$, bottom panel). The empty red histograms are the residual distributions for mass-weighted quantities, while the grey histograms illustrate the residual distributions for light-weighted quantities. We also plot the 0.16, 0.5, and 0.84 quartiles of the residual distributions: in red dashed lines, the quartiles for the mass-weighted residuals, in black, the quartiles for the light-weighted. The quartiles are reported at the top of the panels, in red for the MW residuals, and in grey for the LW ones. The figure shows the following:

(i) The residual distribution for the mass-weighted metallicity is characterized by $q_{16} = -0.31$, $q_{50} = -0.07$, and $q_{84} = 0.13$, and $q_{16} = -0.34$, $q_{50} = -0.12$, and $q_{84} = 0.07$ for light-weighted metallicity.

(ii) The residual distribution for the mass-weighted age is characterized by $q_{16} = -0.18$, $q_{50} = 0.04$, and $q_{84} = 0.26$, and $q_{16} = -0.17$, $q_{50} = 0.06$, and $q_{84} = 0.32$ for light-weighted age.

These results show that our analysis does not introduce any biases: the residuals are well consistent with zero within the 68 per cent confidence intervals. As to be expected, the realistic observational effects implemented in iMaNGA lead to a scatter in the residuals that is well consistent with the measurement errors.

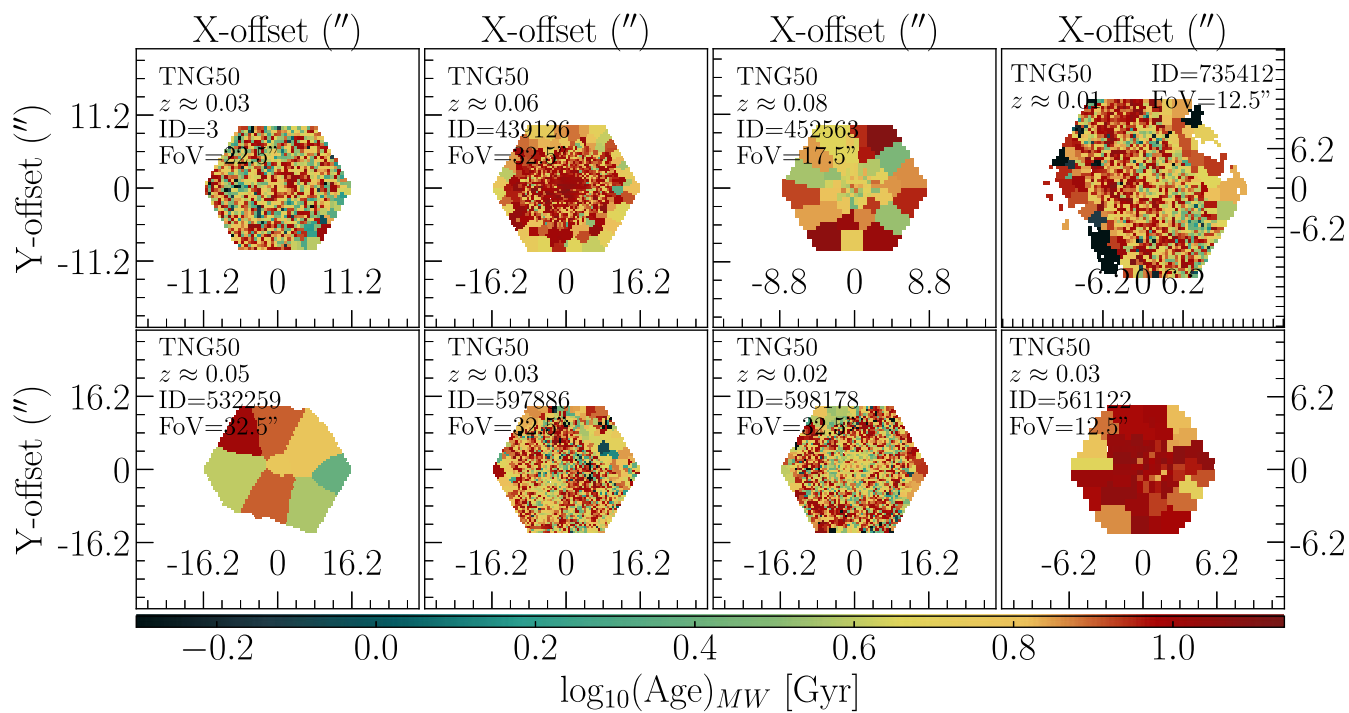

**Figure 20.** As in Fig. 14 for the stellar age, as measured by FIREFLY.

In Appendix A, we further show the residual as a function of the intrinsic properties (see Fig. A1) and the relation between the residuals on metallicity and age (see Fig. A2). We conclude that we are able to recover the intrinsic properties of the TNG50 galaxy sample.

*5.5.2 Star formation histories*

The SFH is an essential quantity to study galaxy formation and evolution processes, but it is also one of the most difficult features to retrieve from observational data. We can use the FIREFLY full-spectral fitting algorithm to resolve the SFH non-parametrically so that we can compare it against the intrinsic SFH from TNG50. Again, we take into account both mass-weighted and light-weighted metrics.

Fig. 22 reports the SFH as SSP mass-weights (top left-hand panel) and light-weights (bottom left-hand panel) versus look-back time. In the top panels, we plot the recovered SFH (teal histogram) and the intrinsic SFH (black hatch-filled histogram) for two example galaxies, i.e. galaxy ID 3 (on the left) and ID 561 122 (on the right) at $z \sim 0.03$. The two galaxies have different morphologies (see Table 1) and different stellar population properties (see Figs 19 and 20). The central panels show the intrinsic SFH as empty black histogram and the recovered SFH in yellow. In the bottom panels, we report the residuals, i.e. the difference between what is predicted by FIREFLY and what is intrinsic to the simulations. Light-weighted quantities are yellow, mass-weighted quantities are empty teal histograms. Intrinsic SFHs are computed from all the stellar particles in the simulated galaxies within the FOV of the hexagonal fiber bundle assumed to observe its light; recovered SFHs are obtained considering all the Voronoi tassels in the galaxies, from the analysis with FIREFLY.

A further test is presented in Fig. A3, Appendix A where we present the SFHs of 20 galaxies, divided into ellipticals and spirals. We conclude that the overall shape of the SFH is reproduced well by the full-spectral fitting procedure.

## 6 DISCUSSION

Other works have presented methods to obtain mock MaNGA data from simulations (Ibarra-Medel et al. 2018; Duckworth, Tojeiro & Kraljic 2019; Nevin et al. 2021; Bottrell & Hani 2022; Sarmiento et al. 2023). In the following, we briefly compare these works to our project.

### 6.1 Ibarra-Medel et al. (2018)

Ibarra-Medel et al. (2018) post-process two simulated Milky Way-sized galaxies (from Colín et al. 2016). The synthetic spectra are constructed using a combination of MILES and GRANADA models (more details in Cid Fernandes et al. 2013), with 156 spectral templates that cover 39 stellar ages (from 1 Myr to 14.2 Gyr), 4 metallicity ($Z/Z_\odot = 0.2, 0.4, 1,$ and $1.5$), and a wavelength range between 3600 and 7000 Å. When generating the synthetic spectra, the SSP models are not interpolated to the exact intrinsic quantities as we do, rather the 'closest' models in terms of age and metallicity are adopted. Because of this choice, Ibarra-Medel et al. (2018) have to distinguish between intrinsic and assigned stellar particles' properties. Also different from us, these authors do not run a radiative transfer code to model dust effects, but apply a simple dust extinction model.

The MaNGA fiber-bundle configuration is mimicked: each stellar particle is associated with one of the MaNGA fibers, each with an FoV of 2.5 arcsec, and the final spectrum in each fiber is given by the stack of all the spectra in it. To each stellar particle, one of the spectra from the templates is assigned, therefore, the final spectrum in each fiber is formed by a combination of discrete ages and metallicities.

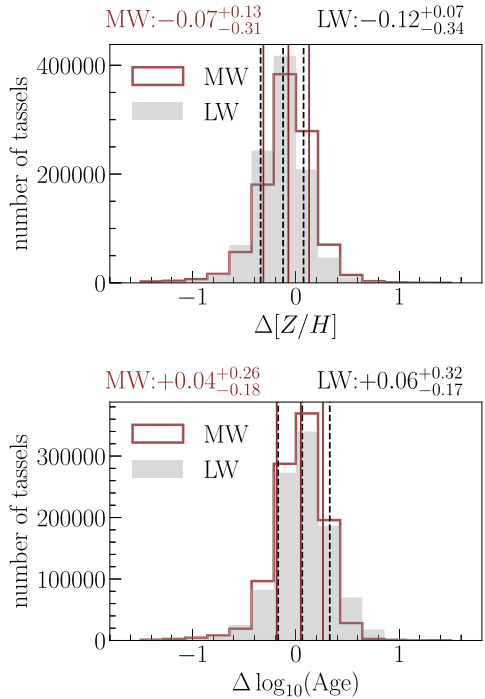

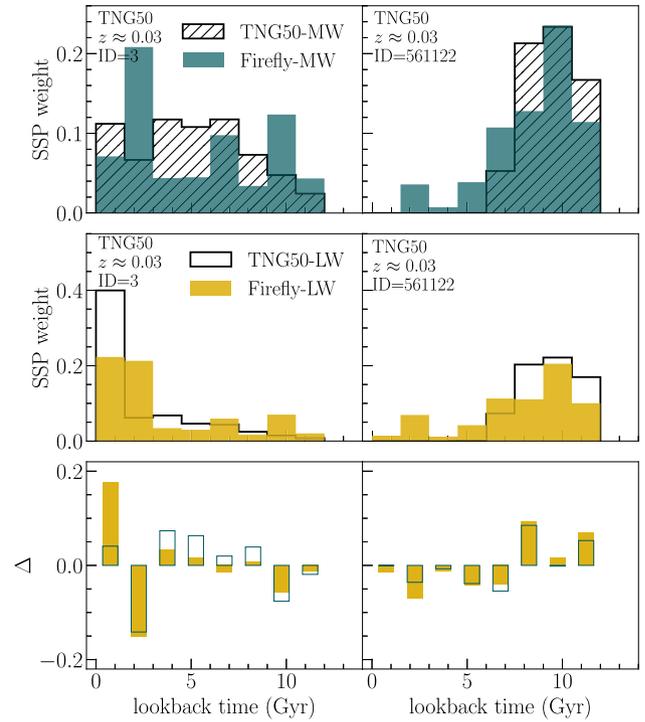

**Figure 21.** Residuals in metallicity and age of the stellar populations as measured by FIREFLY and intrinsic for in the iMaNGA sample. The red empty histograms show the residual distribution for the mass-weighted metallicity ($\Delta[Z/H]_{MW} = [Z/H]_{MW,Firefly} - [Z/H]_{MW,TNG50}$, upper panel) and mass-weighted age ($\log_{10}(Age)_{MW} = \log_{10}(Age)_{MW,Firefly} - \log_{10}(Age)_{MW, TNG50}$, bottom panel); the vertical red lines are the 0.16, 0.5, and 0.84 quartiles of these residual distributions. Grey histograms show the residual distribution for the light-weighted metallicity ($\Delta[Z/H]_{LW} = [Z/H]_{LW,Firefly} - [Z/H]_{LW,TNG50}$, upper panel) and light-weighted age ($\log_{10}(Age)_{LW} = \log_{10}(Age)_{LW,Firefly} - \log_{10}(Age)_{LW, TNG50}$), bottom panel); the vertical black dashed lines are the 0.16, 0.5, and 0.84 quartiles for these residual distributions; quartiles values are reported on top of the panels, in red for the MW residuals distributions, in black for the LW ones.

This operation is repeated three times to mimic the dithering in MaNGA observations. The noise is implemented with a combination of sigmoid functions to mimic the eBoss spectrograph behaviours at short and long wavelengths. This sample of mock galaxies is then used to test the ability of Pipe3D (Sánchez et al. 2016) to recover the stellar populations' properties. Ibarra-Medel et al. (2018) find some biases in the recovered properties, in particular, they recover younger stellar populations in the inner, older regions, and slightly older stellar populations in the outer, younger regions, as well as lower global stellar masses.

### 6.2 Duckworth et al. (2019)

Duckworth et al. (2019) construct the spectra for around 4500 galaxies in TNG100 (Marinacci et al. 2018; Naiman et al. 2018; Nelson et al. 2018; Pillepich et al. 2018b; Springel et al. 2018) adopting the FSPS stellar population models (Conroy & Gunn 2010). FSPS contains 22 metallicity steps (from $\log_{10}Z = -3.7$ to $\log_{10}Z = -1.5$), and 94 age steps (from $\log_{10}t = -3.5$ [Gyr] to $\log_{10}t = 1.15$ [Gyr]), and a wavelength rage between 3600 and 7000 Å. They bin particles within spaxels of a length-side of 0.5 arcsec, over a hexagonal fibre bundle footprint, mimicking the MaNGA datacubes. In each bin, the mean velocity, velocity dispersion, and total flux for all particles are computed. No extinction is assumed. The SNR as a function of the effective radius in the *g* band is computed and used to assign noise to the synthetic spectra. A Gaussian kernel with 2 arcsec FWHM is assumed to mimic the MaNGA PSF. The datacubes are then re-binned with the Voronoi algorithm as in the official MaNGA DAP. Then, the intrinsic properties of the stellar and gas particles in each Voronoi tassel are used to investigate the relationship between the rotation of stars and gas with morphology and halo spin. The analysis is conducted over both TNG100 galaxies and MaNGA galaxies within the paper. They find a good agreement between TNG100 and MaNGA.

**Figure 22.** The SFH recovered by FIREFLY and intrinsic to TNG50, for two example galaxies in iMaNGA. The SFHs are represented as SSP MW (upper panels) or LW (central panels) as a function of lookback time. *Left-hand panels*: the MW SFH of galaxy ID 3 at $z \sim 0.03$. In the top panel, the teal histogram shows the SFHs resolved by FIREFLY compared to the black hatch-filled histogram which illustrates the 'intrinsic' MW SFH reconstructed from TNG50 stellar particle data. *In the central panel*, as the upper one, this time considering the light-weighted (LW) SFH. In particular, the LW SFH as recovered by FIREFLY (yellow histogram), and intrinsic to TNG50 galaxies (black empty histogram). In the bottom panels we report the residuals, i.e. the difference between what is recovered tanks to FIREFLY and what is intrinsic to the simulations; with the empty teal histograms we report the residuals for the MW SSP weights and in yellow for the LW ones. *Right-hand panels*: as the left ones, with the same meaning of symbols, this time for galaxy ID 561 122 at $z \sim 0.03$.

### 6.3 Nevin et al. (2021)

Nevin et al. (2021) test an algorithm to identify galaxy mergers with stellar kinematics. Five merging galaxies and matched isolated ones in GADGET (Springel, Yoshida & White 2001a) are considered. The stellar populations are modelled with the code SUNRISE (Jonsson 2006) based on STARBURST99 stellar population synthesis models (Leitherer et al. 1999), and synthetic datacubes are created. STARBURST99 has 5 metallicities (between $\log_{10}Z = -1.40$ and $\log_{10}Z = -3$.), and the age coverage is between 1 Myr and 1 Gyr, with a wavelength range between 0.009 and 160 μm. Also in this work, as in ours, and differently from Ibarra-Medel et al. (2018), the stellar spectra are interpolated and therefore, the intrinsic stellar population properties are used to define each spectra. The effects of dust attenuation as well as AGN are included. A Gaussian kernel with FWHM equal to 2.5 arcsec is used to mimic the PSF in MaNGA. Then, the datacubes are re-binned to have a spatial sampling of 0.5 arcsec. Also in this work, the morphology is studied with STATMORPH on *SDSS*-like *r*-band images. The hexagonal FoV of the smallest fiber bundles in MaNGA capable of observing the galaxies within $1.5R_{\rm eff}$ is adopted. They produce a typical noise spectrum, which is then normalized and used to include random noise to each spaxel in the datacube. The MaNGA DAP is followed, except that pPXF is run only for the stellar component.

The methodology of this work is most similar to ours, albeit with some important differences. The convolution with the PSF in our pipeline is the last step, after the inclusion of the noise, such that the noise between adjacent spaxels correlates. Also, different from Nevin et al. (2021), we do not assume a Gaussian kernel for the PSF, but we reconstruct a MaNGA effective PSF, which includes dithering effects and seeing. The PSF in our approach is wavelength dependent like the ePSF in MaNGA. Instead of reconstructing a typical MaNGA noise, we reconstruct SNR as a function of the wavelength at $1.5R_{\rm eff}$ and then this information is used as in equation (4) in Paper I. Finally, we adopt population model spectra based on stellar spectra observed with the same *SDSS* spectrograph used for the MaNGA galaxies.

### 6.4 Bottrell & Hani (2022)

Bottrell & Hani (2022) present the code REALSIM-IFS, which is capable of modelling the instrumental sampling mechanics for any IFS instrument, MaNGA included. REALSIM-IFS includes the effects of atmospheric seeing, IFU fibre characteristics and set-up (designs), dithered exposure strategy, line-spread function, and spatial reconstruction of fibre measurements. Instead, Bottrell & Hani (2022) test the application of REALSIM-IFS with TNG50 to create synthetic MaNGA stellar kinematic maps. To our knowledge, this method is the only one capable of reconstructing all the details of the MaNGA observational setup. On the other hand, spectra are not included in their work.

### 6.5 Sermiento et al. (2023)

Sarmiento et al. (2023) apply the method by Ibarra-Medel et al. (2018) to TNG50 galaxies, using template spectra from Bruzual (private communication) based on the MaStar stellar library (Yan et al. 2019). These include a total of 273 synthetic spectra, with 39 ages between 0.0023 and 13.5 Gyr and 7 metallicities between 0.0001 and 0.43, and a linearly sampled wavelength range between 2000 and 10 000 Å. Stellar particles are associated with the closest on among these 273 synthetic spectra as in Ibarra-Medel et al. (2018). This implies that also in thier case, and differently from us, 'assigned' properties do not necessarily coincide with the 'intrinsic' properties predicted by the simulations (see their figs 6 and 7). The difference between assigned and intrinsic properties jeopardises the comparison with the recovered values, leading to larger scatters and potentially larger residuals.

### 6.6 Mock MaNGA catalogues

Among these works, Duckworth et al. (2019) and Sarmiento et al. (2023) produced mock MaNGA catalogues. In both, the approach was to associate a galaxy in TNG with an observed MaNGA galaxy directly. Duckworth et al. (2019) look for unique matches for a total of 4500 galaxies in TNG100. Galaxies are matched by stellar mass, size, and *SDSS g − r* colour. Sarmiento et al. (2023) follow a similar approach, matching galaxies in mass, redshift, and effective radius, but they do not match galaxies uniquely rather allow galaxies drawn from TNG50 to be selected multiple times. When a TNG50 galaxy is selected multiple times, it is observed with different line of sights. In this way, Sarmiento et al. (2023) match the full MaNGA catalogue to galaxies in TNG50, that are however not physically distinct.

Their approach is different from ours. In our work, we adopt the TNG50 catalogue and apply the MaNGA selection criteria to it, in order to construct our mock MaNGA sample. As a consequence, there are no direct matches between MaNGA and TNG50 galaxies. Because we wanted to match the MaNGA observed sample selection criteria, our simulated iMaNGA catalogue is significantly smaller than the MaNGA sample because of the relatively small simulated volume in TNG50.

Both approaches have benefits and pitfalls, but they appear to serve different scopes. In our approach, we aim to observe the universe as generated by the Illustris simulation as if it was observed, rather than matching the simulation to the observation. As a consequence, our iMaNGA catalogue allows us to test the characteristics of the simulation, such as fraction of certain galaxy types, galaxy scaling relations, etc. For example, we find a paucity of massive elliptical galaxies. Interestingly, residual biases are still observed by Sarmiento et al. (2023), including an excess of disc galaxies at high mass in agreement with our findings.

## 7 CONCLUSIONS

We present a mock MaNGA catalogue, called iMaNGA, generated from the state-of-the-art magnetohydrodynamical galaxy simulations TNG50. We illustrate the general characteristics of the iMaNGA sample in terms of morphology, kinematics, and stellar populations. We further run detailed tests comparing intrinsic galaxy properties from TNG50 with the recovered properties in iMaNGA derived through full-spectral fitting.

We identify galaxies in TNG50 through friends-of-friends algorithms, and initially select by redshift ($0.01 \leq z \leq 0.15$) and number of stellar particles ($N > 10\,000$), and exclude all galaxies that are flagged as of non-cosmological origin (for more details see Genel et al. 2017; Pillepich et al. 2018b, and the Data Specification page for IllustrisTNG)[7]. These criteria lead to the selection of 48 248 galaxies. Since we want to recover a smooth distribution in spatial sampling, as for the MaNGA-Primary sample, we then randomize galaxy redshifts in the initial sample, and select galaxies by *i*-band magnitude and redshift. In this way, we directly mimic the MaNGA selection criteria, which yields our 'parent sample' containing 3152 galaxies. We finally impose a flat distribution in absolute *i*-band magnitude following the MaNGA survey design, by randomly selecting unique galaxies from the parent sample. The resulting sample comprises 1000 galaxies from TNG50 and represents the final iMaNGA catalogue. We then post-process our iMaNGA galaxies following the method presented in Paper I (Nanni et al. 2022). We investigate whether the iMaNGA sample reproduces general trends of the MaNGA-Primary Sample. We find that angular sizes and spatial resolution (both in kpc and in terms of $R_{\rm eff}$) are well consistent between iMaNGA and MaNGA. Likewise, the relationship between angular size and total stellar mass is matched well. The correlations of morphology with angular size and stellar mass are instead not recovered by the iMaNGA sample. These differences are driven by the fact that TNG50 is dominated by late-type systems with a paucity of intermediate-mass lenticular galaxies and massive elliptical galaxies as also found in e.g. Huertas-Company et al. (2019).

---

[7]https://www.tng-project.org/data/docs/specifications/

We demonstrated a generally good agreement between 'recovered' and 'intrinsic' properties, including stellar kinematics, stellar population ages, metallicities, and star formation histories. In particular, the stellar kinematics is recovered running the full-spectral fitting code PPXF, as in the MaNGA DAP. The 'intrinsic' stellar peculiar velocity and the stellar velocity dispersion along the LOS are both recovered within $1\sigma$. The stellar populations' properties are recovered by running another full-spectral fitting code, i.e. FIREFLY. Both the 'intrinsic' stellar age and stellar metallicity are well recovered, i.e. the residuals over all tassels in the iMaNGA sample are consistent with zero within the 68 per cent confidence interval. We also show the 'recovered' and 'intrinsic' SFHs for 22 galaxies in the iMaNGA catalogue, which show a generally good agreement.

Finally, we present a comparison with other works which have produced mock MaNGA datacubes and/or mock MaNGA catalogues.

While there are a number of differences in methodology for the construction of the mock MaNGA data cube, the most important difference to highlight for this paper is the method of constructing the mock MaNGA sample. Other works in the literature assign simulated galaxies from TNG50 or TNG100 directly to MaNGA galaxies by matching their basic properties such as mass and effective radius, whereas we take the simulations at face value and apply the target selection criteria from MaNGA to the simulated catalogue. Both approaches have merits. While the former aims at producing a theoretical catalogue that resembles MaNGA as closely as possible, our approach is designed to test galaxy formation models. In our next paper in this series, we will present the scientific analysis of the iMaNGA catalogue in direct comparison with MaNGA.

## ACKNOWLEDGEMENTS

The authors would like to thank Chris Lovell and Harry Desmond for their feedback during the revision process. LN is supported by an STFC studentship. STFC is acknowledged for support through the Consolidated Grant Cosmology and Astrophysics at Portsmouth, ST/S000550/1. Numerical computations were done on the Sciama High Performance Compute (HPC) cluster which is supported by the ICG, SEPnet, and the University of Portsmouth. Funding for the *Sloan Digital Sky Survey* IV has been provided by the Alfred P. Sloan Foundation, the U.S. Department of Energy Office of Science, and the Participating Institutions. SDSS-IV acknowledges support and resources from the Center for High Performance Computing at the University of Utah. The *SDSS* website is www.sdss.org. SDSS-IV is managed by the Astrophysical Research Consortium for the Participating Institutions of the SDSS Collaboration including the Brazilian Participation Group, the Carnegie Institution for Science, Carnegie Mellon University, Center for Astrophysics | Harvard & Smithsonian, the Chilean Participation Group, the French Participation Group, Instituto de Astrofísica de Canarias, The Johns Hopkins University, Kavli Institute for the Physics and Mathematics of the Universe (IPMU)/University of Tokyo, the Korean Participation Group, Lawrence Berkeley National Laboratory, Leibniz Institut für Astrophysik Potsdam (AIP), Max-Planck-Institut für Astronomie (MPIA Heidelberg), Max-Planck-Institut für Astrophysik (MPA Garching), Max-Planck-Institut für Extraterrestrische Physik (MPE), National Astronomical Observatories of China, New Mexico State University, New York University, University of Notre Dame, Observatário Nacional/MCTI, The Ohio State University, Pennsylvania State University, Shanghai Astronomical Observatory, United Kingdom Participation Group, Universidad Nacional Autónoma de México, University of Arizona, University of Colorado Boulder, University of Oxford, University of Portsmouth, University of Utah, University of Virginia, University of Washington, University of Wisconsin, Vanderbilt University, and Yale University. The primary TNG simulations were realised with compute time granted by the Gauss Centre for Supercomputing (GCS): TNG50 under GCS Large-Scale Project GCS-DWAR (2016; PIs Nelson/Pillepich), and TNG100 and TNG300 under GCS-ILLU (2014; PI Springel) on the GCS share of the supercomputer Hazel Hen at the High Performance Computing Center Stuttgart (HLRS).

## DATA AVAILABILITY

The iMaNGA catalogue is made available through the following website: http://www.icg.port.ac.uk/imanga/. A list of all the galaxies post-processed and analysed in this paper, i.e. the iMaNGA sample, is available here: https://github.com/lonanni/iMaNGA/. Finally, the iMaStar code can be found here: https://github.com/lonanni/iMaNGA.

MaNGA data are part of SDSS-IV, publicly available at https://www.sdss4.org/dr16/manga/manga-data/ (Abdurro'uf et al. 2022). The FIREFLY code is available at: https://www.icg.port.ac.uk/FIREFLY and the MaStar population models at https://www.icg.port.ac.uk/mastar. Illustris and IllustrisTNG data are publicly available at https://www.illustris-project.org/data (Nelson et al. 2019a).

## APPENDIX A: RECOVERY OF STELLAR POPULATION PROPERTIES

In this appendix, we present additional tests of our capability to recover stellar population properties age and metallicity. Fig. A1 shows the residuals as a function of the intrinsic properties, for both mass-weighted and light-weighted quantities, colour coded by the number of tassels. As in Fig. 21, the residuals are consistent with 0 within the $1\sigma$ confidence level for the majority of the tassels. Small, negative correlations are present for age, indicating slightly positive residuals (hence overestimation of age) at low ages and slightly negative residuals (hence underestimation of age) at old ages. No such bias is seen for metallicity.

Fig. A2 shows the relation between the residuals in metallicity and age, for both mass-weighted (left-hand panel) and light-weighted (right-hand panel) quantities, colour coded by the number of tassels. We can see again that the residuals are consistent with zero within the $1\sigma$ confidence level (the green solids lines report the 0 value). The degeneracy between age and metallicity is visible for the light-weighted quantities, i.e. when metallicity is overpredicted, age is underpredicted and vice versa.

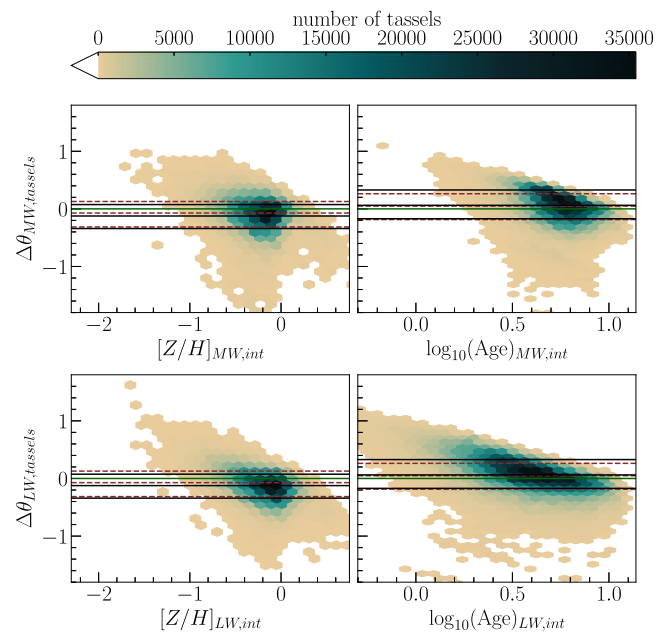

**Figure A1.** The distribution of the residuals of the metallicity (left-hand panels) and age (right-hand panels) of the stellar populations as measured by FIREFLY and by TNG50, for all the Voronoi tassels in the iMaNGA sample, as a function of the intrinsic values, colour coded by the number of tassels, considering mass-weighted (top panels) and light-weighted (bottom panels) results. The horizontal lines are the 0.16, 0.5, and 0.84 quartiles of these residual distributions, both light-weighted (red dashed lines) and mass-weighted (black solid lines). The green solid lines illustrate the 0.

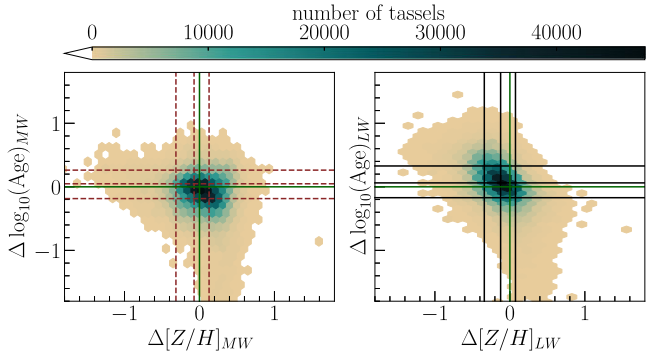

**Figure A2.** The distributions of the residuals of the stellar populations' age as a function of the residuals in the stellar populations' metallicity as measured by FIREFLY and by TNG50, for all the Voronoi tassels in the iMaNGA sample, as a function of the intrinsic values, colour coded by the number of tassels, considering mass-weighted (left-hand panel) and light-weighted (right-hand panel). The horizontal lines are the 0.16, 0.5, and 0.84 quantiles of these residual distributions, both light-weighted (red dashed lines) and mass-weighted (black solid lines). The green solid lines report the zero-values.

Finally, we discuss the SFH recovered by FIREFLY considering all the Voronoi bins for each galaxy, following the same method as in Section 5.5.2. In Fig. A3, we plot 20 'intrinsic' SFHs in comparison with the 'recovered' SFHs for galaxies in a redshift range between 0.03 and 0.1, and a total stellar mass between 9.5 and 11. $\log_{10} M_*/M_\odot$. We separate between ellipticals (for a Sérsix index above 2.5, left-hand columns) and spirals (for a Sérsix index between 0.5 and 1, right-hand columns). Some systematic differences between intrinsic and recovered SFHs become apparent. The recovered SFHs tend to be flatter (more uniform in time) than the intrinsic SFHs, which is more pronounced for the MW SFHs. Our fitting procedure seems to add a fraction of old populations, which is then balanced by a younger component.

For the spirals, instead, a preferred population at around 2 Gyr seen in the recovered SFHs. A more detailed analysis of these discrepancies in light of our methodology for spectral fitting will be subject of further work, which goes beyond the scope of this paper. It is important to underline here that for the 'intrinsic' SFH history, all the stellar population particles present in the simulated galaxies are considered, while the 'recovered' SFH is instead obtained running FIREFLY over the Voronoi-tessellated galaxy.

The galaxies in this mass and redshift range categorized as ellipticals are (listed here as snapshot-id in TNG50): 96–216478, 97–279747, 97–22, 96–480963, 95–496556, 96–8, 96–98927, 97–568905, 96–395465, 96–5, 95–477720, 96–585516 and 96, 535026.

The spiral galaxies are: 95–181049, 92–489917, 96–528258, 95–514781, 95–215985, 95–499904, 93–119273, 93–483018, 97–388073, 92–490816, 94–481051 and 95–493913.

This paper has been typeset from a T<sub>E</sub>X/LAT<sub>E</sub>X file prepared by the author.

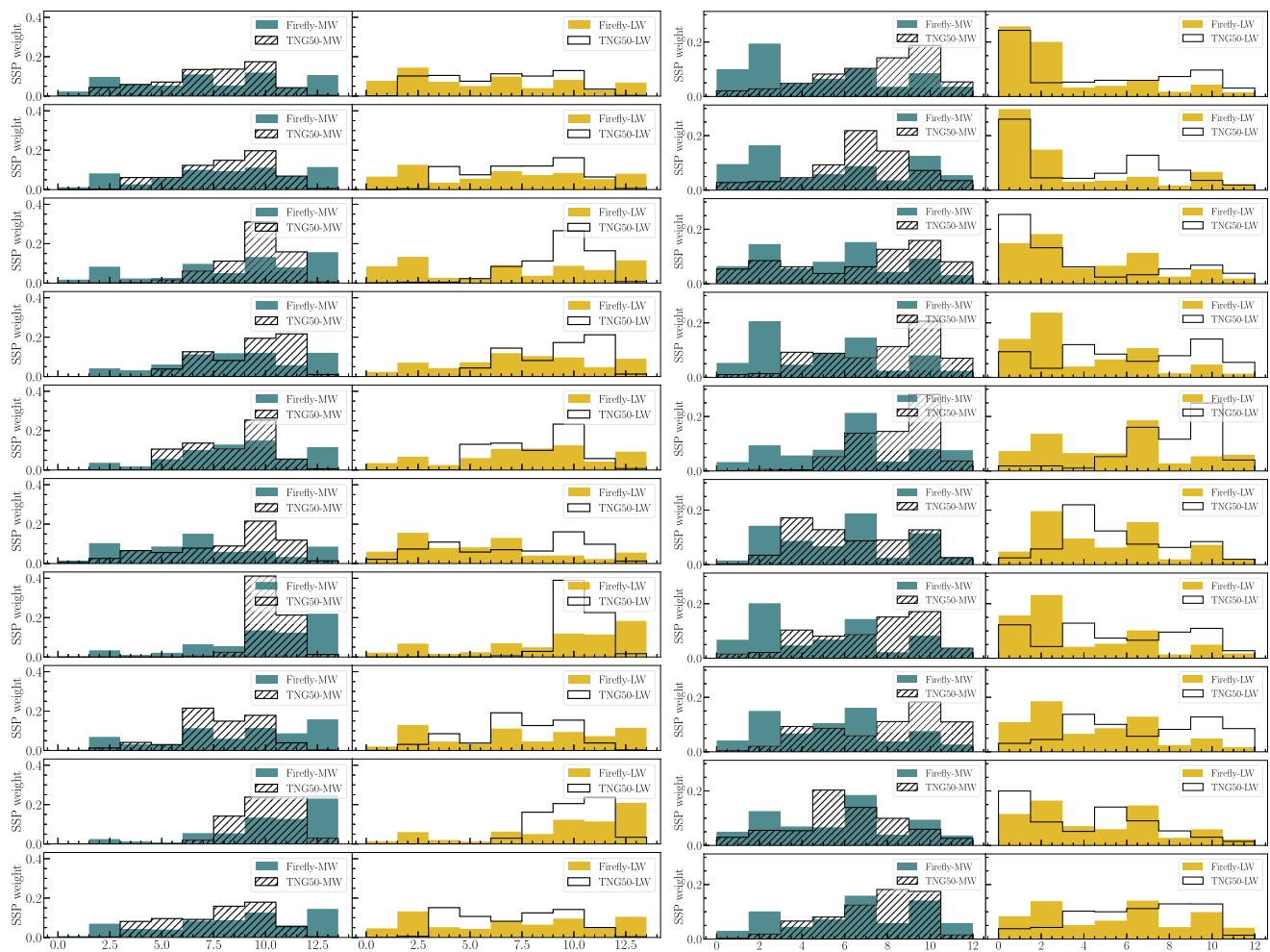

**Figure A3.** The SFH recovered by FIREFLY and computed from the stellar particle information provided by TNG50, for 20 galaxies in the iMaNGA sample. These galaxies are selected in the iMaNGA sample with a redshift range between 0.03 and 0.1, and a total stellar mass between 9.5 and 11. $\log_{10}M_*/M_\odot$, and dividing them into ellipticals (when the Sérsix index is above 2.5) and spirals (when the Sérsix index is between 0.5 and 1). *First two columns*: the SFH of elliptical galaxies in the defined redshift and mass range. In the first column, the teal histograms show the SFHs resolved by FIREFLY compared to the black hatch-filled histograms which illustrates the 'intrinsic' SFH reconstructed from TNG50 stellar particle data. Here, the SFHs are represented as SSP mass-weights as a function of lookback time. The second column, as the first one, this time considering the SFHs as SSP light-weights as a function of lookback time. In particular, the SFHs as recovered by FIREFLY (yellow histograms), and 'intrinsic' to TNG50 galaxies (black empty histograms). *Last two columns*: as the first two columns, this time for spiral galaxies in the discussed range in mass and redshift.